\documentclass[a4paper,fleqn,usenatbib]{mnras}

\pdfoutput=1

\usepackage{txfonts}

\usepackage[T1]{fontenc}
\usepackage{ae,aecompl}

\usepackage{graphicx}	
\usepackage{amsmathMC}	
\usepackage{amssymb}	
\usepackage[flushleft]{threeparttable} 
\usepackage{float}

\title[Bending waves and halo substructure]{Bending Waves in the Milky Way's disc from halo substructure}

\author[M. H. Chequers et al.]
{Matthew H. Chequers,\thanks{E-mail: 12mhc@queensu.ca} Lawrence M. Widrow\thanks{E-mail: widrow@queensu.ca} and Keir Darling
\\
Department of Physics, Engineering Physics \& Astronomy, Queen's
University, Kingston, ON K7L 3N6, Canada}

\date{Accepted XXX. Received YYY; in original form ZZZ}

\pubyear{2018}

\begin{document}
\label{firstpage}
\pagerange{\pageref{firstpage}--\pageref{lastpage}}
\maketitle


\begin{abstract}

 \noindent We use $N$-body simulations to investigate the excitation
 of bending waves in a Milky Way-like disc-bulge-halo system.  The
 dark matter halo consists of a smooth component and a population of
 subhaloes while the disc is composed of thin and thick components.
 Also considered is a control simulation where all of the halo mass is
 smoothly distributed.  We find that bending waves are more vigorously
 excited in the thin disc than the thick one and that they are
 strongest in the outer regions of the disc, especially at late times.
 By way of a Fourier decomposition, we find that the
 complicated pattern of bending across the disc can be described as a
 superposition of waves, which concentrate along two branches in
 the radius-rotational frequency plane. These branches correspond to vertical
 resonance curves as predicted by a WKB analysis.  Bending waves in
 the simulation with substructure have a higher amplitude than those
 in the smooth-halo simulation, though the frequency-radius
 characteristics of the waves in the two simulations are very similar.
 A cross correlation analysis of vertical displacement and bulk
 vertical velocity suggests that the waves oscillate largely as simple
 plane waves.  We suggest that the wave-like features in astrometric
 surveys such as the Second Data Release from \textit{Gaia} may
 be due to long-lived waves of a dynamically active
 disc rather than, or in addition to, perturbations from a recent
 satellite-disc encounter.
\end{abstract}





\section{Introduction}
\label{sec:intro}

The Milky Way's disc bends in and out of its midplane.  The most
conspicuous example of this is the large scale warp, which has been
observed in H\,{\sc i} \citep{levine2006}, dust
\citep{freudenreich1994}, and stars \citep{djorgovski1989}.  It starts
within the Solar Circle and increases in amplitude toward the edge of
the disc \citep{drimmel2001,lopez-corredoira2002,momany2006,
  reyle2009,schonrich2017}.  There is also evidence for
short-wavelength ripples or corrugations in the disc
\citep{xu2015,schonrich2017} as well as a mix of localized bending and
breathing motions in the Solar Neighbourhood \citep[][but also see
  \citealt{sun2015}, \citealt{ferguson2017}, \citealt{pearl2017},
  \citealt{carrillo2018}, and \citealt{wang2018} for more recent
  developments]{widrow2012,carlin2013,williams2013}.  Further evidence
for perturbations perpendicular to the disc midplane comes from
North-South asymmetries in stellar number counts \citep{widrow2012,
  yanny2013,ferguson2017}.  Finally, wave-like features have been seen
in the Gaia Data Release 2 (GDR2) \citep{gaiadr2disckin2018,
  antoja2018, poggio2018}.  In particular, \citet{antoja2018} found
spiral-like structures in the $\{z,\,v_R,\,v_\phi,\,v_z\}$ phase space
distribution of stars within a circular arc at a Galactocentric radius
just beyond the Solar Circle.

In this paper, we focus on the Milky Way's stellar disc and adopt the
view that the aforementioned phenomena can be understood as a
superposition of bending waves.  On the observational front, the
\textit{Gaia} space observatory \citep[][but see \citealt{gaiadr22018}
  for an overview of GDR2]{perryman2001,gaiamission2016} will
allow us to determine the bulk vertical velocity, $V_z$, and midplane
displacement, $Z$, as a function of position in the disc plane across
a substantial fraction of the disc \cite[see][]{gaiadr2disckin2018}.
In principle, one might then try to decompose $Z$ and $V_z$ into waves
that are defined by their azimuthal wave number, radial wavelength,
and pattern speed.  As with spiral structure, one might further
attempt to determine whether bending waves are leading or trailing and
whether they are prograde or retrograde with respect to the Galactic
rotation.

Our aim in this paper is to study the physics of bending waves through
$N$-body simulations of an isolated disc-bulge-halo system.  Our work
differs from previous simulation-based studies in several important
respects.  First, we follow the evolution of a model galaxy when a
fraction of the halo mass is replaced by a system of subhaloes.  For
the most part, earlier simulations followed a single satellite as it
plunged through the disc with the goal of connecting observations with
specific disc-subhalo (or disc-satellite galaxy) events
\citep{gomez2013,widrow2014,feldmann2015,
  delavega2015,donghia2016,gomez2016,gomez2017,laporte2018}. Indeed,
some of the observed phenomena have been attributed to the Sagittarius
dwarf galaxy and/or the Large Magellanic cloud
\citep{gomez2013,laporte2017}.

We take the somewhat different viewpoint that the disc is continually
perturbed by subhaloes and satellite galaxies.  Local perturbations
are sheared into circular arcs that then oscillate as self-gravitating
waves.  The key idea is that the vertical perturbations observed today
may come more from long-lived waves than very recent
disc-substructure interactions.

\citet{delavega2015} have suggested that the evolution of
perturbations in the disc can be understood as purely kinematical.  To
make their case, they follow test-particle representions of disc
perturbations in a time-independent unperturbed potential.  Their
perturbations therefore phase-mix but do not self-gravitate.  The
resulting structures are found to be qualitatively similar to the
bending and breathing patterns seen in the Solar Neighbourhood.  The
viewpoint that perturbations evolve kinematically forms the basis of
the analysis by \citet{antoja2018} who use phase-mixing arguments to
`date' the event that produced the phase space features they found in
GDR2. However, the inconsistent timescales between various phase-wrapping models (for example, \citealt{minchev2009}, \citealt{antoja2018}, and \citealt{monari2018}) suggests the situation is more complex in that other dynamical processes, such as the bar or spiral structure \citep{kawata2018,quillen2018}, may also play a significant role in driving the evolution of features seen in the phase space distribution of stars.

Our view is that immediately after the disc encounters a satellite the
resulting vertical perturbation may well evolve kinematically, but that once
sheared into an extended arc, it will behave as a self-gravitating
wave.  This picture is supported by the good agreement between the
results from linear perturbation theory, which includes self-gravity
\citep{hunter1969, sparke1984, sparke1988, nelson1995}, and
simulations \citep{chequers2017}.  Eventually, the energy imparted to
the disc by a passing subhalo or satellite is converted into random
motions of the disc stars thereby vertically heating the disc
\citep{lacey1985,toth1992,sellwood1998}.  Thus, the life-cycle of a
perturbation involves excitation, phase-wrapping or shearing across
the disc, self-gravitating wave-like action, and disc heating.

The challenge then is to see if one can disentangle these different
processes.  Adding to the complexity of the problem is the realization
that vertical bending waves are excited even in simulations where, apart from the shot noise of the $N$-body distribution, the
halo is smooth and axisymmetric \citep{chequers2017}.  In what follows, we make
direct comparisons between a control simulation, where the halo is
smooth, and one run with a clumpy halo.  Our set-up closely follows
that of \citet{gauthier2006} who considered an equilibrium
disc-bulge-halo model for M31, which, when initialized with a smooth
halo, was stable against bar formation for at least 10 Gyr.  However,
when $\sim$~10 per cent of the halo mass was replaced by orbiting
substructure, the disc developed more vigorous spiral structure and a
strong bar (see also \citealt{dubinski2008} and
\citealt{kazantzidis2008}).  Not surprisingly, halo substructure also
excites vertical oscillations of the disc.  We find that the amplitude
of bending waves, including the warp, increases by a factor $\sim 5$
relative to the waves that arise in the smooth halo run.

Vertical perturbations can also be excited by internal mechanisms,
such as a bar \citep{monari2015} and spiral structure
\citep{debattista2014,faure2014,monari2016a,monari2016b}.  Indeed, one
expects that any time-dependent perturbing potential that sweeps
through the disc will cause it to contract and expand in the direction
perpendicular to the midplane.  Bending waves require a perturbation
that breaks symmetry about the midplane, as would occur with a
buckling bar.  Since the focus of this paper is on the effects of
substructure, we consider a relatively low mass disc that is stable to
bar formation for at least 10 Gyr.  We stress that our goal is not to develop the most
realistic model for the Milky Way but rather to study the physics of
bending waves that are generated by halo substructure.

A second feature of our simulations is the inclusion of both thin and
thick disc components.  We use the Action-based Galaxy Modelling
Architecture \citep[\textsc{agama},][]{vasiliev2018} code to generate
initial conditions.  Action-based methods do not rely on the epicycle
approximation and are therefore particularly well-suited for models
that include a thick disc component. One motivation for including two
components in our model disc is the conjecture that dynamically
distinct components will respond differently to a passing subhalo
\citep{banik2017}.  In particular, we expect that the transfer of
energy between a subhalo and disc is enhanced when the timescale for
changes in the gravitational potential match (are in resonance with)
the vertical epicyclic motions of the disc stars \citep{sellwood1998}.
Moreover, slow moving subhaloes tend to excite bending motions while
fast moving ones excite compression and rarefaction.  Since the
vertical frequency for thick disc stars is larger than for thin disc
stars, one might expect a preference for bending waves in the thin
disc and breathing waves in the thick disc \citep{widrow2014}.

As mentioned above, one might hope to use \textit{Gaia} data to
characterize bending waves in the Milky Way's stellar disc according
to properties such as wavelength and pattern speed.  Ultimately, one
might imagine decomposing disc perturbations into something akin to
normal modes.  The theoretical problem of determining the normal modes
of a self-gravitating stellar dynamical disc is challenging even in
highly idealized cases.  \citet{mathur1990}, \citet{weinberg1991}, and
\citet{widrow2015}, for example, worked out the vertical oscillations
of a self-gravitating, isothermal sheet \citep{spitzer1942,camm1950}.
Though these one-dimensional modes can be long-lived with respect to
the timescale for the vertical epicyclic motions of the stars, they
eventually decay due to phase mixing and Landau damping.
\citet{toomre1966} and \citet{araki1985} stressed the difference
between bending and density waves in a self-gravitating
plane-symmetric system.  With density waves, gravity causes overdense
regions to collapse with velocity dispersion (i.e., kinematic
pressure) providing the restoring force.  By contrast, bending waves
are enhanced by the centrifugal force as particles pass through a bend
in the disc.  In this case, gravity acts to restore a perturbed region of the disc to its equilibrium position.

A complementary approach is to ignore the velocity dispersion of the
stars and focus instead on the dynamics of a rotating disc.  In the
pioneering work of \citet{hunter1969}, and subsequent analyses by
\citet{sparke1984} and \citet{sparke1988}, the disc is treated as a
set of concentric rings, which interact with one another
gravitationally.  The normal modes of the `$N$-ring' system can then be found using standard eigenvalue methods.  In the limit of large $N$, the modes form a continuum.  Therefore any initial perturbation, which naturally involve a superposition of these modes, will disperse or phase mix.

With simulations, such as the ones described in this paper, we have
the luxury of knowing the full phase space distribution function (DF),
or at least an $N$-body sample of the DF of the stars, across the
entire disc at all times.  As described below, this allows us to
analyse the perturbations using various mathematical tools.  Here, we
draw on previous studies of spiral density waves and consider
representations of $Z$ and $V_z$ in terms of azimuthal wave number
and/or frequency.  This spectral analysis demonstrates a clear
connection between results from linear theory and WKB analyses, on the
one hand, and fully self-consistent, non-linear simulations on the
other.  In addition, we compute the cross-correlation between $Z$ and
$V_z$ and find a rich structure in time and Galactocentric radius.

The outline of the paper is as follows: In Section~\ref{sec:sims} we
describe the initial conditions and $N$-body parameters for our
simulations.  We focus on subhaloes and their interactions with the
disc in Section~\ref{sec:subhalointeractions}.  In
Sections~\ref{sec:fourierwaveanalysis} and
\ref{sec:spectralwaveanalysis} we apply Fourier methods in the
analysis of these simulations.  We further explore the wave-like
nature of bending waves by comparing midplane displacement and
vertical bulk motion in Section~\ref{sec:wavelikenature}.  We discuss
some implications of our results in Section~\ref{sec:discussion}
before concluding in Section~\ref{sec:conclusions}.

\section{Simulations}
\label{sec:sims}

We simulate the evolution of a disc-bulge-halo model galaxy comprised
of thin and thick disc components as well as a halo where 10 per cent
of the halo mass is initially in subhaloes and the remaining 90 per
cent is smoothly distributed.  We also run a simulation
where the halo is smooth.  A comparison of the two simulations, which
we refer to as the satellite and control/isolated simulations, allows us to
quantify the effects of disc-subhalo interactions.

\subsection{An action-based equilibrium model for the Galaxy}
\label{subsec:actionbased}

We construct an equilibrium model for a Milky Way-like galaxy using
the action-based
code\footnote{https://github.com/GalacticDynamics-Oxford/Agama.  The
  version of \textsc{agama} we used pre-dates the official public
  release and was downloaded on July 27, 2017.} \textsc{agama}
\citep{vasiliev2018}.  The code employs an iterative scheme to find
self-consistent forms for the DF of the different components and the
gravitational potential.  The general idea is to begin with an initial
guess for the potential and expressions for the DFs in terms of a set
of three integrals of motion.  One then computes the density as a
function of coordinates and solves Poisson's equation to obtain a new
approximation for the potential.  The procedure is repeated until the
density-potential pair converges.  In our previous work, we used the
\textsc{galactics} code where the DFs were chosen to be functions of
the angular momentum about the symmetry axis of the galaxy, the total
energy, and the vertical energy \citep{kuijken1995,widrow2008}.  The
latter is only approximately conserved in thin discs and not at all
well conserved in thick ones.  An action-based code avoids this
problem since the actions are identically conserved.  The price one
pays is the computational challenge of computing the actions. \textsc{agama} uses an efficient and accurate implementation of the so-called `St{\"a}ckel fudge' \citep{binney2012} to compute the action transformations for all galaxy components \cite[][but see also \citealt{sanders2016} for a review of approaches]{vasiliev2018}.

With \textsc{agama}, the input parameters explicitly determine the
functional form of the DFs in action space and only implicitly
determine the galaxy's structure.  Thus, a certain
amount of trial and error is required to construct a model with the
desired properties.  In this section, we outline the properties of our
`target' model.  The most important \textsc{agama} input file
parameters are given in Table~\ref{tab:initialmodelparams}.

We assume that the density profiles of the bulge and halo
are given, at least approximately, by the Hernquist
\citep{hernquist1990} and NFW \citep{navarro1996} profiles,
respectively.  Both of these components can be modelled within
\textsc{agama} by the double-power DF from \citet{posti2015}, which
yields a system whose spherical density profile is given, to a good
approximation, by
\begin{equation}
\rho (r) = \frac{\rho_0}{(r/r_b)^\alpha (1 + r/r_b)^{\beta - \alpha}}~.
\label{eq:doublepower}
\end{equation}
The disc components in \textsc{agama} are modelled using the
pseudo-isothermal DF from \citet{binney2011}.  In constructing these
models one uses the fact that though the galactocentric radius $R$ of
a star is not an integral of motion, its angular momentum $L_z$ is.
One can therefore use $R_c$, the radius of a circular orbit with
angular momentum $L_z$, as a proxy for $R$.  For example, the factor
in the disc DF that controls the vertical structure has the form
\begin{equation}
f\left (L_z,\,J_z\right ) = \frac{\nu_z}{\left (2\pi\sigma_z^2\right )^{1/2}}
e^{-\nu_z J_z/\sigma_z^2}~,
\end{equation}
\noindent where $\nu_z$ is the vertical epicycle frequency and $J_z$
the vertical action.  The quantity $\sigma_z$ is a function of $R_c$
(i.e., a function of $L_z$) and controls the radial profiles of the
vertical scale height and velocity dispersion.  In our models,
$\sigma_z$ is assumed to be an exponentially descreasing function of
$R_c$, which yields a disc with an approximately exponential vertical
velocity dispersion profile.

\begin{figure} 
\includegraphics[width=\columnwidth]{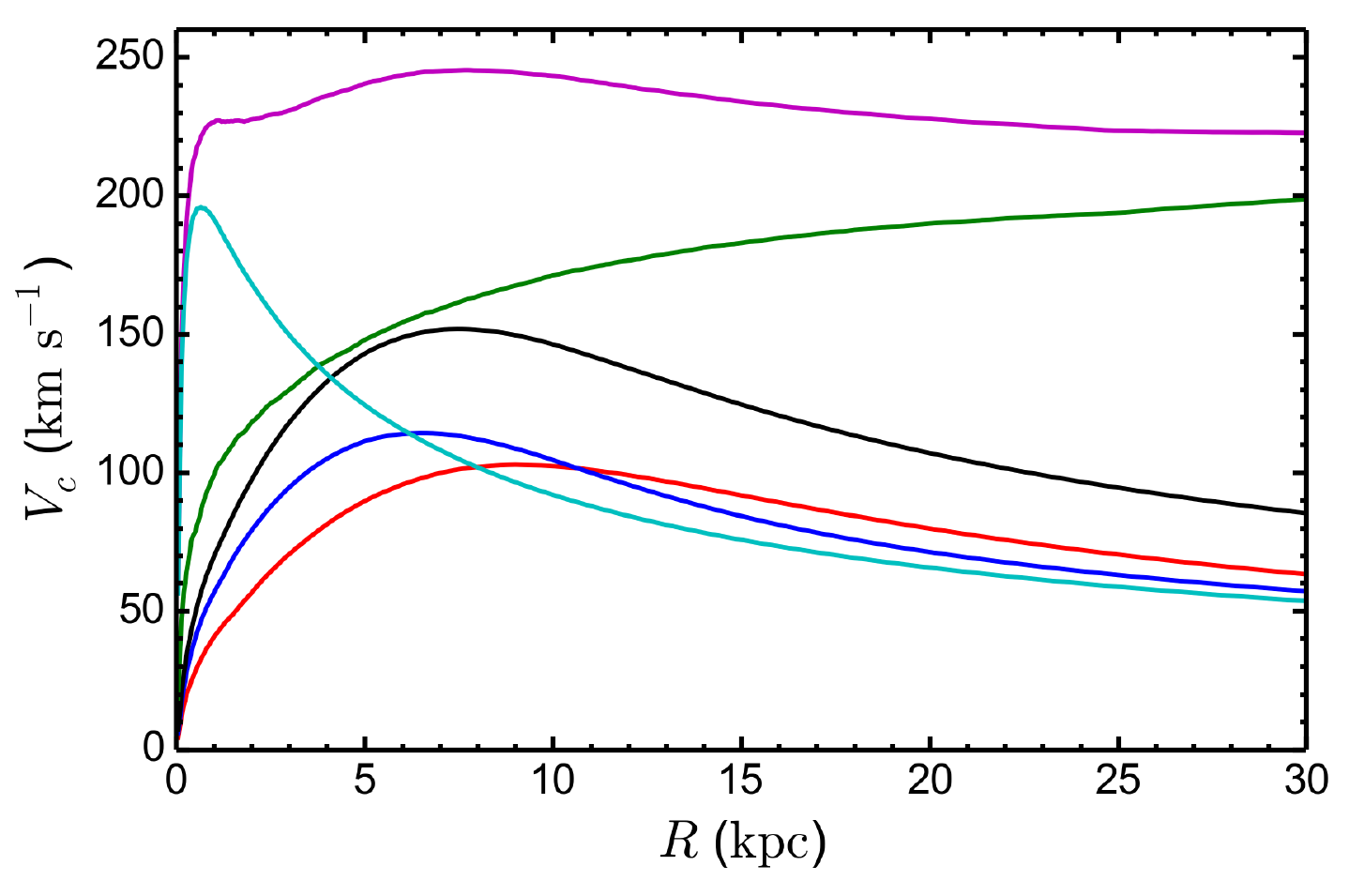}
\caption{Circular speed curve decomposition for our Milky Way-like
  model. The solid magenta curve shows the total circular speed of the
  model. Also shown are the contributions to the circular speed curve
  from the bulge (cyan), halo (green), thin disc (blue),
  thick disc (red), and total disc (black).
\label{fig:rotcurve}}
\end{figure}

Our choice of disc parameters is guided by the analysis of
\citet{bovy2012b} and \citet{bovy2013} as well as our desire for a
model that is stable against the formation of a bar for the duration of the simulation runtime.  As discussed in
Section~\ref{sec:intro}, this model allows us to focus on the
effect substructure, rather than a bar or large-amplitude spiral
structure, has on vertical oscillations. The thin disc has a radial exponential scale length of $3.6\,{\rm kpc}$ and a mass of $2.6\times 10^{10}\,M_\odot$. The corresponding quantities for the thick disc are $2.0\,{\rm kpc}$ and $2.2\times 10^{10}\,M_\odot$. The total disc mass of $4.8\times 10^{10}\,M_\odot$ is in accord with the total stellar disc mass for the Milky Way reported in the literature (\citealt{bovy2013}, but see also \citealt{courteau2014} for a review). The vertical structure of the discs are well-described by a ${\rm sech}^2 (z/2 \, z_d)$ profile with $z_d$ = $260\,{\rm pc}$ for the thin disc and $690\,{\rm pc}$ for the thick \citep{bovy2012b}.

Our bulge has a target density profile given by the Hernquist profile
(equation~\ref{eq:doublepower} with $\left (\alpha,\,\beta\right ) = \left
(1,\,4\right )$) and a total mass of $2 \times 10^{10}~M_{\odot}$,
which is slightly higher (by about 5-10 per cent) than the current best
estimates for the Milky Way's bulge \citep[see][]{bland-hawthorn2016}.
The bulge scale radius is taken to be $600\,{\rm pc}$ as in the 
\citet{bovy2013} model and the velocity distribution is approximately 
isotropic.

The halo target density profile is given by the NFW profile
(equation~\ref{eq:doublepower} with $\left (\alpha,\,\beta\right ) = \left
(1,\,3\right )$; see \citealt{navarro1996}) with scale radius $r_b=22\,{\rm kpc}$ and
$\rho_0 = 9\times 10^6 M_\odot\,{\rm kpc}^{-3}$.  The parameters for
the DF are chosen so as to yield a total circular speed curve that is
approximately flat with $V_c \simeq 220-250\,{\rm km\,s^{-1}}$ for
$2\,{\rm kpc} < R < 30\,{\rm kpc}$ (see Fig.~\ref{fig:rotcurve}).  Note
as well that the total disc is submaximal.  In particular,
$V_{d}^{2}/V_{c}^{2} \approx 0.37$ at the peak of the total disc's
circular speed curve ($R \sim$7~kpc), where $V_d$ is the contribution to the
circular speed from the total disc.

\begin{figure} 
\includegraphics[width=\columnwidth]{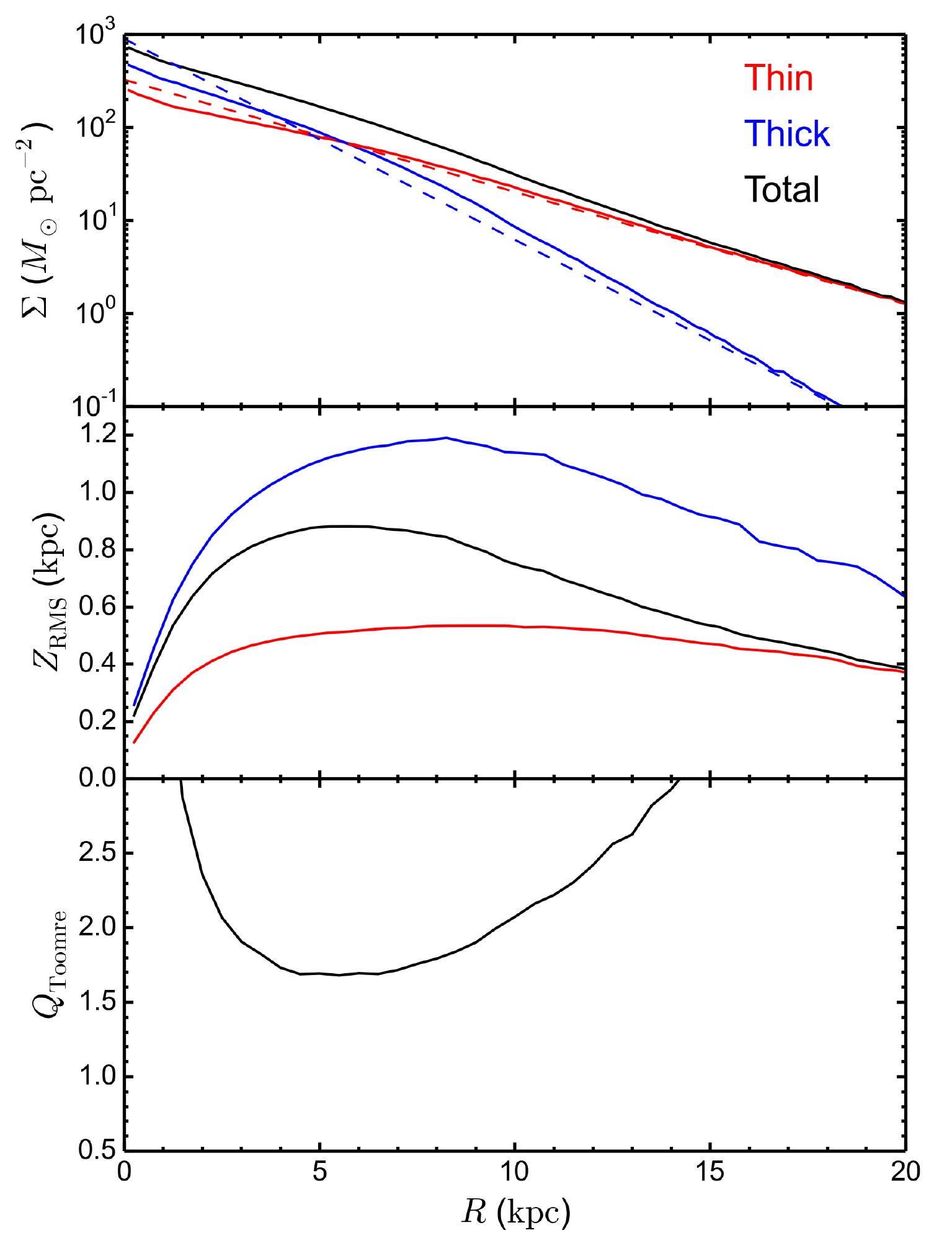}
\caption{Radial properties of our Milky Way-like disc. The colour
  corresponds to the thin (red), thick (blue), and total (black)
  discs. Top panel: Surface density profiles. Solid lines correspond
  to our $N$-body model and dashed lines indicate the target
  profiles. Middle panel: Root-mean-square height profiles. Bottom
  panel: Toomre $Q$ parameter profile of the total disc.
\label{fig:discICs_radial}}
\end{figure}

Fig.~\ref{fig:discICs_radial} shows the surface density profiles of
the thin and thick discs. We note that, at small radii, the surface
densities are supressed relative to a pure exponential.  This effect
is likely due to the presence of a cuspy bulge and the fact that in
action-based modelling the resulting density law is only approximately
equal to the target. Our Hernquist bulge causes the epicyclic
frequencies to rapidly increase as $R\to 0$ and consequently the
pseudo-isothermal DF is not well behaved in the central disc
\citep{vasiliev2018}.

The root-mean-square thicknesses of the discs in our model vary with $R$
by a factor of $\sim 2$ as seen in the middle panel of
Fig.~\ref{fig:discICs_radial}.  The depression at small radii is,
again, due to the inclusion of a cuspy bulge. The decrease in
thickness at larger radii is presumably due to using an exponential
vertical velocity dispersion profile, which would produce a constant
scale height only for a thin disc in isolation.

The central radial dispersion was chosen to yield a relatively high
Toomre $Q$ parameter \citep{toomre1964} across the disc, as can been
seen in the bottom panel of Fig.~\ref{fig:discICs_radial}.  The fact
that we have a submaximal disc with $Q > 1.5$ indicates that very little
structure will develop in the disc, at least in the absence of 
external perturbations.

\begin{figure} 
\includegraphics[width=\columnwidth]{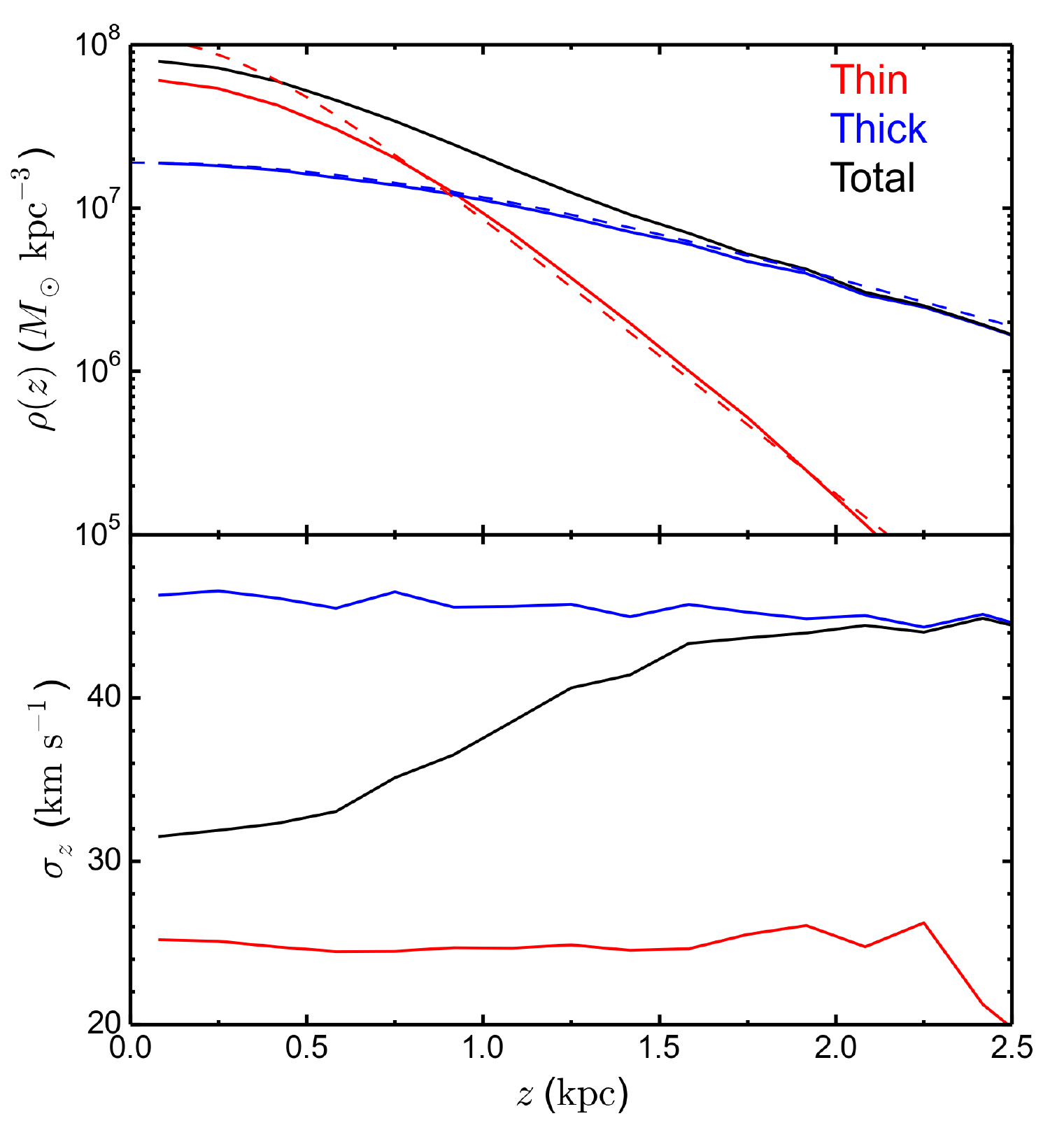}
\caption{Vertical properties of our Milky Way-like disc evaluated at a
  galactocentric radius of 8 kpc. The colour corresponds to the thin
  (red), thick (blue), and total (black) discs. Top panel: Vertical
  density profiles. Solid lines correspond to our $N$-body model and
  dashed lines indicate the target profiles. Bottom panel: Vertical
  velocity dispersion as a function of distance from the midplane.
\label{fig:discICs_vertical}}
\end{figure}

In the top panel of Fig.~\ref{fig:discICs_vertical} we show the
vertical density profiles for the thin, thick, and total discs
evaluated at $R = 8 \, {\rm kpc}$.  On average, the density profiles
roughly correspond to our target parameters. In the bottom panel we
show the vertical velocity dispersion profile, and see that the thin
and thick discs are roughly isothermal. The vertical profiles are
shown for the Solar circle though the same general trends are observed
at all other radii.

\subsection{Halo substructure}

For our satellite simulation we replace 10 per cent of the halo mass with
100 subhaloes by a scheme similar to the one presented in
\citet{gauthier2006}. We randomly remove 10 per cent of the particles
that characterize the smooth halo and insert an equivalent amount of
mass in the form of spherical subhaloes.  These subhaloes have initial
center-of-mass positions and velocities drawn from the same DF as the
original halo.

\begin{figure} 
\includegraphics[width=\columnwidth]{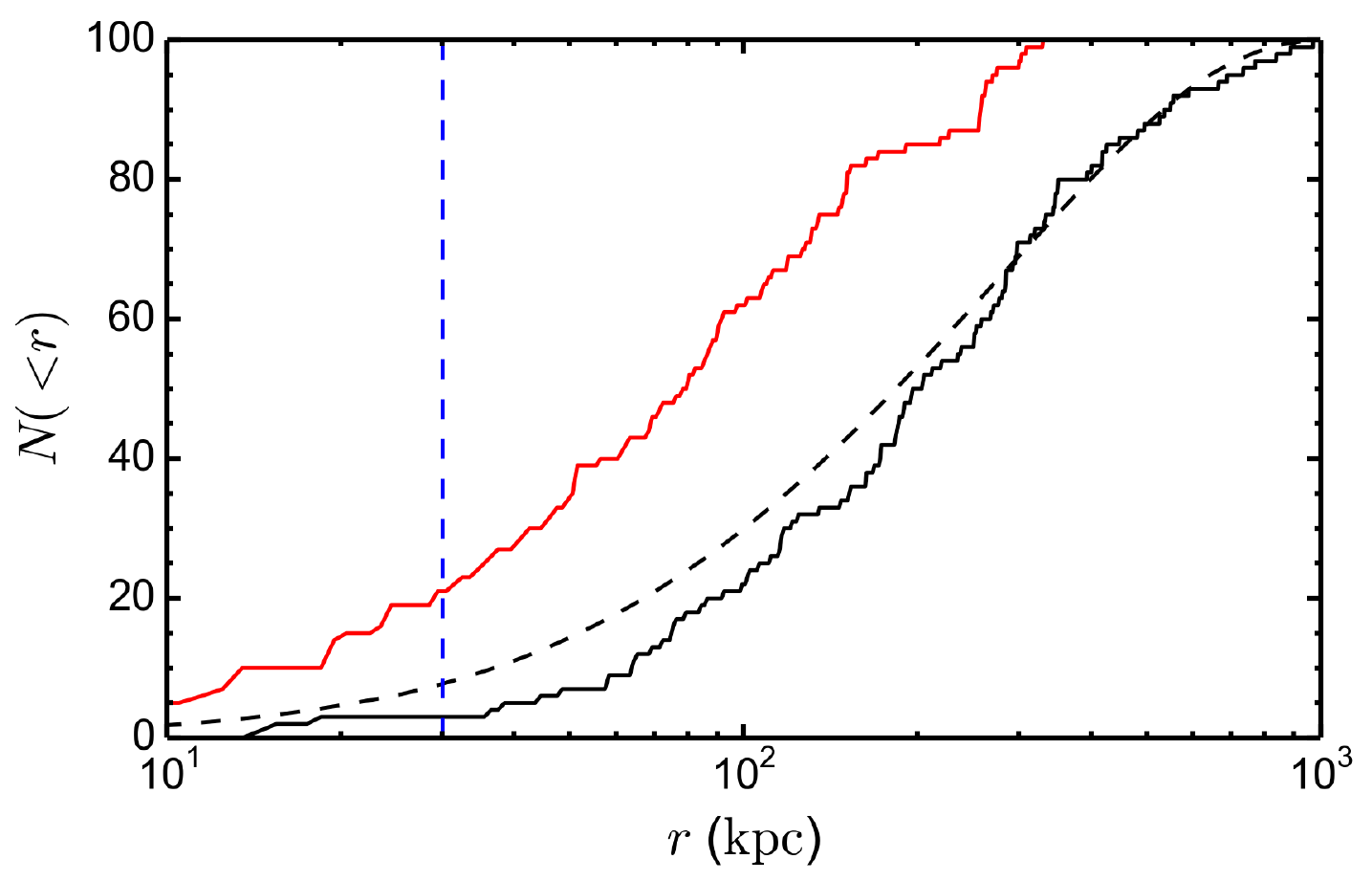}
\caption{Cumulative number distribution of initial center-of-mass subhalo radius within the host halo (solid black). The dashed black curve is the normalized cumulative number distribution of particles in the smooth halo from which the positions of the subhalo system were sampled. The solid red line shows the cumulative distribution of initial pericentre distances of the subhalo population. The vertical dashed blue line references the initial edge of the disc. 
\label{fig:subhaloespherrdistribution}}
\end{figure}

The initial cumulative distribution of subhaloes as a function of
galactocentric radius is shown in
Fig.~\ref{fig:subhaloespherrdistribution}.  Note that while only a few
subhaloes begin with an initial radius inside the region of the disc
(i.e., $r \la 30\,{\rm kpc}$), $20$ or so subhaloes follow orbits with a
perigalacticon less than $30\,{\rm kpc}$.  Thus, most of the subhaloes
that pass through the disc were initialized well outside the disc
region.

\begin{figure} 
\includegraphics[width=\columnwidth]{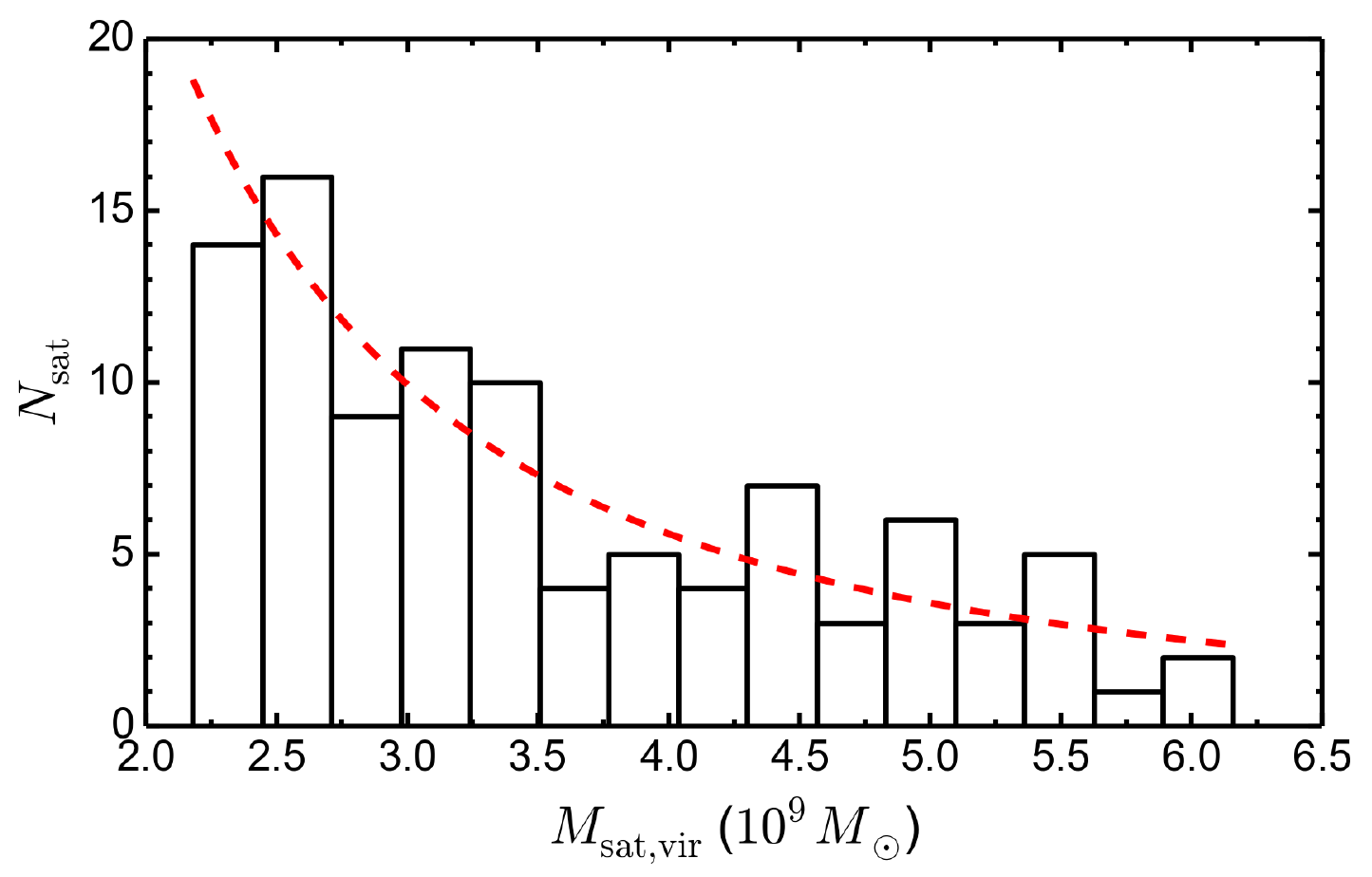}
\caption{Number distribution of substructure virial mass. Solid black
  represents the distribution of the substructure population in our
  subhalo model. The dashed red curve corresponds to the number
  distribution derived from the mass spectrum from which our
  population was sampled (equation~\ref{eq:subhalomasses}).
\label{fig:subhalomassfuncs}}
\end{figure}

We assume an initial subhalo mass function given by
\begin{equation} \label{eq:subhalomasses}
\frac{dN_{\rm sat}}{dM_{\rm sat}} ~=~ 
\begin{cases}
A M_{\rm sat}^{-2}   &  \text{for}~M_{\rm min} \le M_{\rm sat} \le M_{\rm max}~, \\
0 & \text{otherwise}~,
\end{cases}
\end{equation}
\noindent where $A$ is a normalization constant. In addition, we assume that the
total mass in subhaloes is one tenth of the virial mass of the halo
and the maximum subhalo mass is $M_{\rm max} = 0.1 \left (M_{\rm disc} +
M_{\rm bulge} \right )$.  These two conditions allow us to determine $M_{\rm
  min}$ and $A$.  Equation~(\ref{eq:subhalomasses}) is then sampled to
determine initial virial masses for the subhaloes.  We use \textsc{galactics}
\citep{kuijken1995,widrow2008} to generate $N$-body realizations of
the subhaloes assuming that each has a truncated NFW density
profile. \textsc{galactics} requires the specification of NFW mass in
addition to scale and truncation radii. We compute scale radii for
each subhalo as $c = r_{200}/a_s$, where $c$ is the halo
concentration, $r_{200}$ is the radius in which the mean halo density
is 200 times the critical density, and $a_s$ is the NFW scale
radius.  We adopt a constant value of $c = 20$ for all of the
subhaloes, which is in agreement with cosmological simulations for the
mass range of our subhalo population over a significant look-back time
($\sim$10 Gyr) \citep{maccio2007,zhao2009}.  A histogram of the
subhaloes as a function of their virial mass is shown in
Fig.~\ref{fig:subhalomassfuncs}.

We smoothly truncate the $N$-body subhaloes at their tidal radii which
are numerically computed using the Jacobi approximation \citep[section
  8.3]{binney2008}
\begin{equation} \label{eq:jacobi}
\bar{\rho}_{\rm sat} (r_t) = 3 \bar{\rho}_{\rm halo} (r_{\rm sat}),
\end{equation}
\noindent where $\bar{\rho}_{\rm sat} (r_t)$ is the subhalo's mean
density inside the truncation radius $r_t$, and
$\bar{\rho}_{\rm halo}$ is the mean density of the halo inside the initial
radius, $r_{\rm sat}$, of the subhalo.  Truncation is
meant to take into account tidal stripping that may have occurred
prior to the initial epoch of the simulation.

\subsection{\boldmath{$N$}-body simulations}

Multidimensional sampling routines built into \textsc{agama} were used to
generate an $N$-body realization of our model galaxy with $2.5M$ thin
disc, $2.5M$ thick disc, $250k$ bulge, and $5M$ halo particles. When
subhaloes were included, the smooth component of the halo had $4.5M$
particles while each subhalo had $25k$ particles.
The initial conditions were evolved for $\sim 10\,{\rm Gyr}$ using
\textsc{gadget-3} \citep{springel2005} with softening lengths of $40$,
$15$, $100$, and $80$~pc for the disc, bulge, halo, and subhalo
particles, respectively. The maximum time step was set to $0.18\,{\rm
  Myr}$, which is $\sim$0.6 per cent of the galactic dynamical time
defined at the radius of peak disc contribution to the rotation curve,
$R \sim 7$~kpc. Total energy was conserved to within 0.06 per cent. The ratio of total kinetic to potential energy for the particle distributions comprising the thin and thick discs, bulge, and halo in the isolated model was $-0.503$, $-0.502$, $-0.505$, and $-0.503$, respectively, indicating that the initial conditions were sufficiently close to equilibrium. Indeed, in the first 250 Myr we observed a maximum increase in the vertical velocity dispersion and $Z_{RMS}$ profiles of only $\sim 1 \, {\rm km}\,{\rm s}^{-1} $ and $\sim 60 \, {\rm pc}$, respectively.

Over time the discs in our simulations rotate and drift about the
global coordinate origin, particularly in the simulation with halo
substructure.  To account for this when analyzing our models we center
and rotate all model components by the following scheme: We first
iteratively compute the center of mass of disc particles within a
cylinder of radius $20\,{\rm kpc}$ and height $2\,{\rm kpc}$.  Next,
we reorient the system so that the disc lies in the $x$-$y$ plane by
using a two-dimensional Newton-Raphson scheme to find the orientation
that minimizes the root mean square vertical displacement of disc
particles within the same cylinder used to find the centre of mass.

\section{Disc-Subhalo interactions}
\label{sec:subhalointeractions}

In this section we focus on the evolution of the subhalo population in
our satellite simulation.  We define the position and velocity of a
particular subhalo as the average position and velocity of its 100
initially most energetically bound particles.  This method does a good
job at accurately tracking the phase space coordinates of a subhalo
until it is completely disrupted.  Computation of the bound mass
within each subhalo has historically involved an iterative procedure
concerning energy calculations \citep[e.g.][]{benson2004}. Here, we
employ a simpler approach where the bound mass of each subhalo at each
time output is taken to be the mass contained within the Jacobi sphere
as defined in equation~(\ref{eq:jacobi}).

\begin{figure} 
\includegraphics[width=\columnwidth]{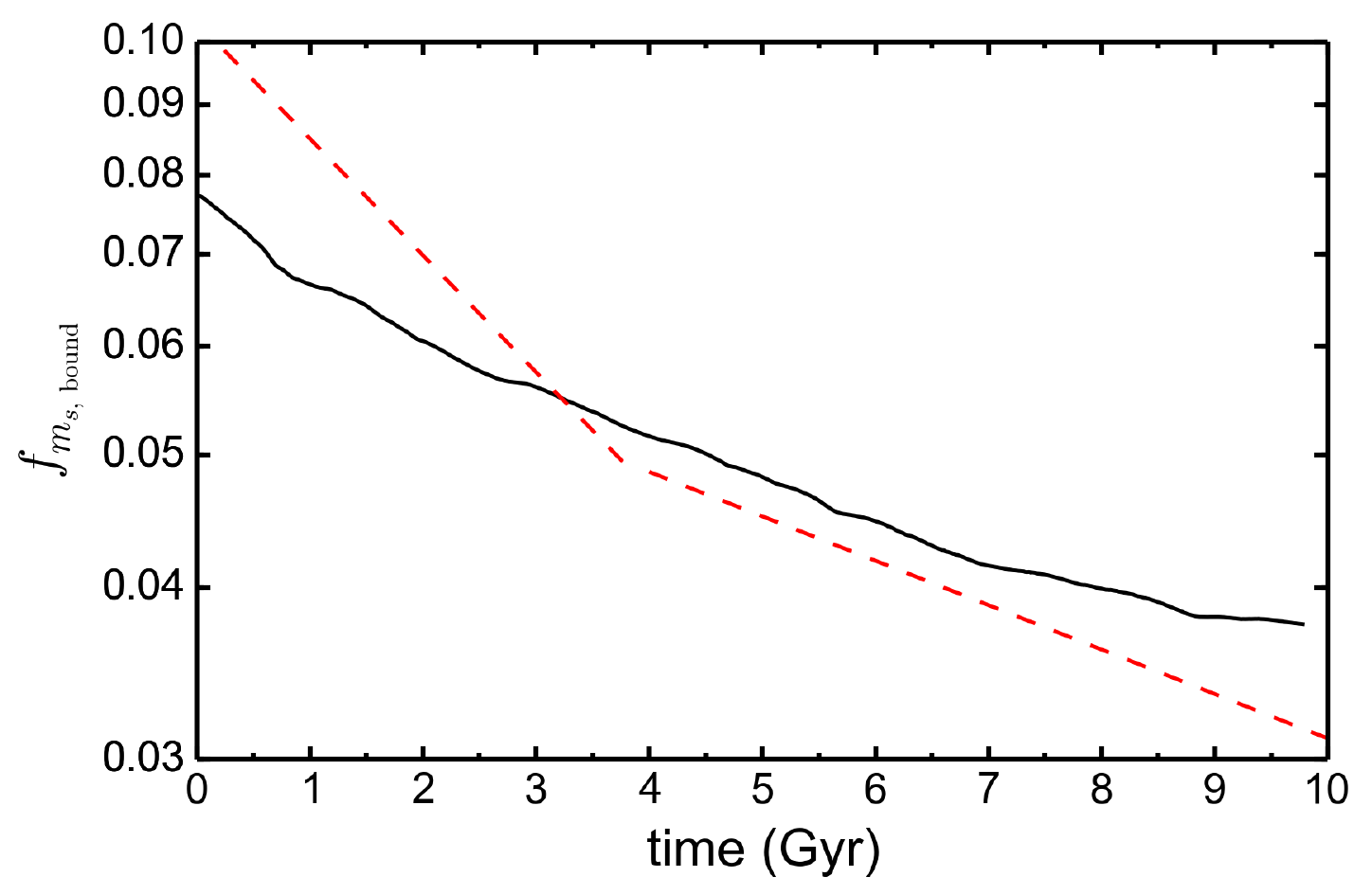}
\caption{ Mass bound to subhaloes relative to the total mass contained in the halo and subhalo population as a function of time (solid black). For comparison we also show the corresponding relation for the two distinct phases of subhalo mass loss in the \citet{gauthier2006} simulation (dashed red).
\label{fig:satenclosedmassvstime}}
\end{figure}

In Fig.~\ref{fig:satenclosedmassvstime} we show the temporal evolution of the mass bound in subhaloes relative to the total mass of the halo and subhalo population. Our initial total substructure mass fraction, $\sim 0.08$, is in rough agreement with that of Milky Way-like haloes found in cosmological simulations, such as the Aquarius Project \citep[][figure 12]{springel2008}. If we consider only subhaloes initiated within a radius of 50 kpc, the substructure mass fraction decreases to $\sim 10^{-2}$. This value also agrees with that found by the Aquarius Project, although in the case of our simulation is due to the stochastic nature of sampling the initial positions and masses of only 100 subhaloes. This is supported by Fig. ~\ref{fig:subhaloespherrdistribution}, where the inner galaxy contains fewer subhaloes than expected, and the fact that those subhaloes initiated within a radius of $\sim$~90 kpc have masses from the lower end of the mass spectrum we sampled. 

Our results can also be compared with those
from \citet{gauthier2006} who argued that the time-dependence of the
mass in subhaloes could be described by two distinct phases.  During
the first $4\,{\rm Gyr}$ $f\propto \exp{\left (-t/\tau_f \right )}$ with
$\tau_f \simeq 5.1\,{\rm Gyr}$, while from $4$ to $10\,{\rm Gyr}$
$\tau_ f \simeq 13.4\,{\rm Gyr}$.  We find a similar trend where the
exponential decay rate of subhalo mass descreases with time, though
our initial `rapid decay phase' is shorter.  The difference may be a
simple reflection of the fact that our distribution extends to much
larger radii and therefore a greater fraction of our subhaloes do not
experience tidal disruption.

\begin{figure} 
\includegraphics[width=\columnwidth]{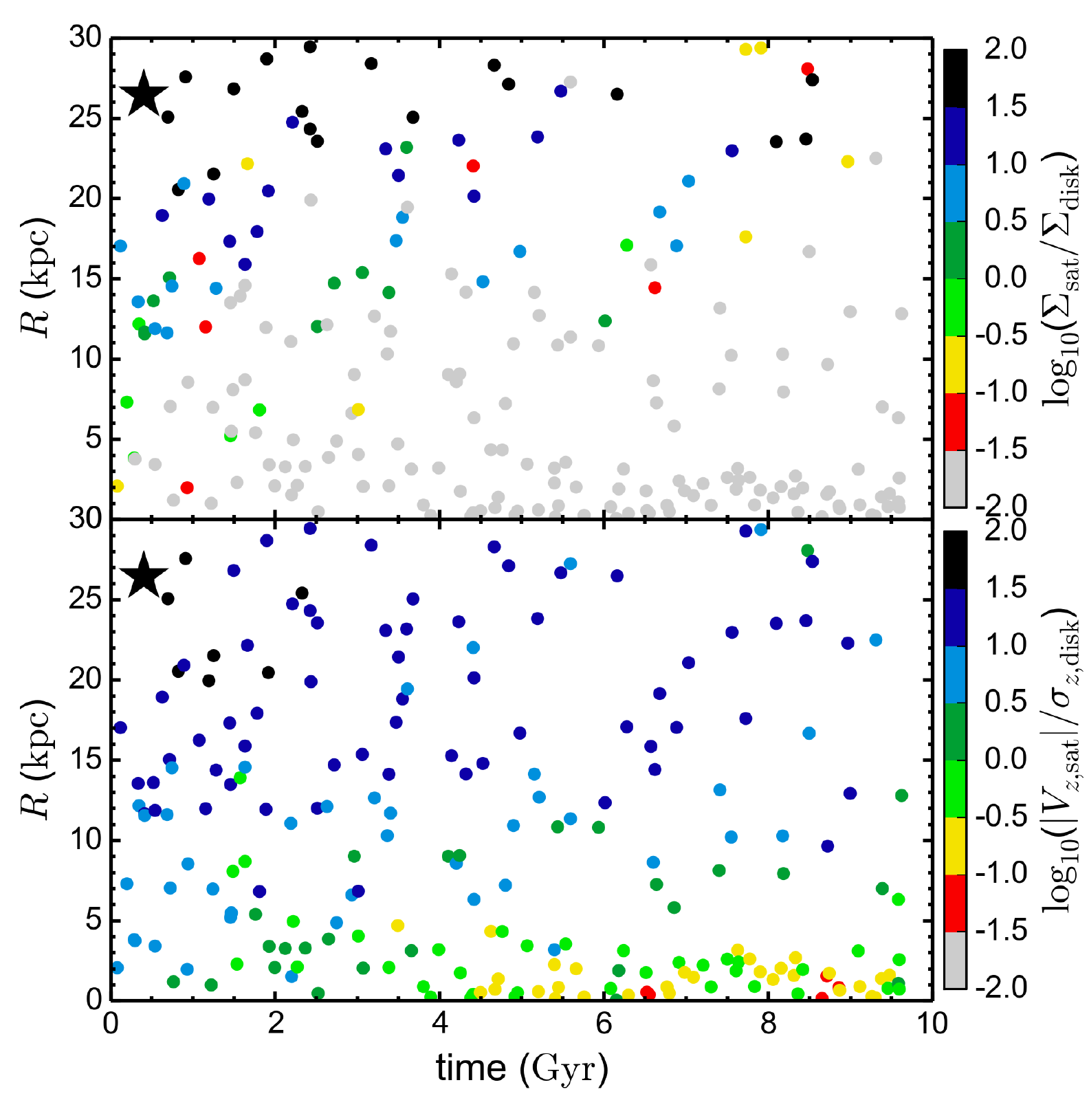}
\caption{Time evolution of subhalo impacts on the disc. Each point
  represents the cylindrical radius and the time at which a subhalo
  crosses the $z = 0$ plane. The star symbol indicates the specific impact event discussed in Section~\ref{sec:fourierwaveanalysis}. Colour coding in the top panel
  corresponds to the ratio of projected subhalo surface density to
  that of the local disc at the time of impact - a proxy for the
  efficacy of each impact to produce vertical waves. In the bottom
  panel, the colour corresponds to the ratio of subhalo vertical
  velocity to the local vertical velocity dispersion of the disc at
  that time. 
\label{fig:satimpact}}
\end{figure}

The amplitude and extent of a perturbation in the disc due to a
passing subhalo will depend on the position at which the subhalo's
orbit crosses the disc plane, its mass and velocity, and the
orientation of the orbital angular momentum vector with respect to
that of the disc's \citep[i.e. prograde, retrograde, or vertical
  orbits;][]{gomez2013,widrow2014,feldmann2015}.  The two panels in
Fig.~\ref{fig:satimpact} are meant to give a sense of which subhaloes
will cause the most significant perturbations.  In both panels we show
the times and cylindrical radii at which a subhalo crosses the disc
midplane.  In the top panel we colour the points according to the
ratio of subhalo surface density to the locally
azimuthally averaged surface density of the disc at the time of impact. The restoring force
of the disc is proportional to the disc surface density, so by
comparing this with the corresponding quantity for the subhaloes we
get a sense of how significantly a given interaction might perturb the
vertical structure of the disc.  In general, we see that the subhaloes
that cross the disc plane with $R< 10\,{\rm kpc}$ have relatively low
surface densities, except perhaps during the first few ${\rm Gyr}$.
The number of significant subhalo interactions (interactions where the
subhalo surface density is comparable to or greater than that of the
disc) increases significantly at larger radii, in part because the
density of the disc decreases exponentially, and in part because
subhaloes tend to be disrupted if their orbits take them close to the
galactic centre.

In the bottom panel of Fig.~\ref{fig:satimpact} we show the subhalo's
vertical velocity as it passes through the disc midplane relative to
the local vertical velocity dispersion of the disc as a function of
$R$ and simulation time.  We see that satellites that pass through the
outer disc have speeds 10-100 times higher than the typical vertical
velocities of the disc stars.  In this region, the epicyclic motion
can all but be ignored during a subhalo-disc interaction.  Conversely,
subhaloes passing through the disc at intermediate radii ($5\,{\rm
  kpc} <R<15\,{\rm kpc}$) have vertical velocities more typical of the
vertical velocities in the disc.  This sets up the possibility for
resonant excitation of both vertical bending and breathing waves
\citep{sellwood1998,widrow2014}.

\section{Fourier analysis of density and vertical waves}
\label{sec:fourierwaveanalysis}

\begin{figure*} 
\includegraphics[width=\textwidth]{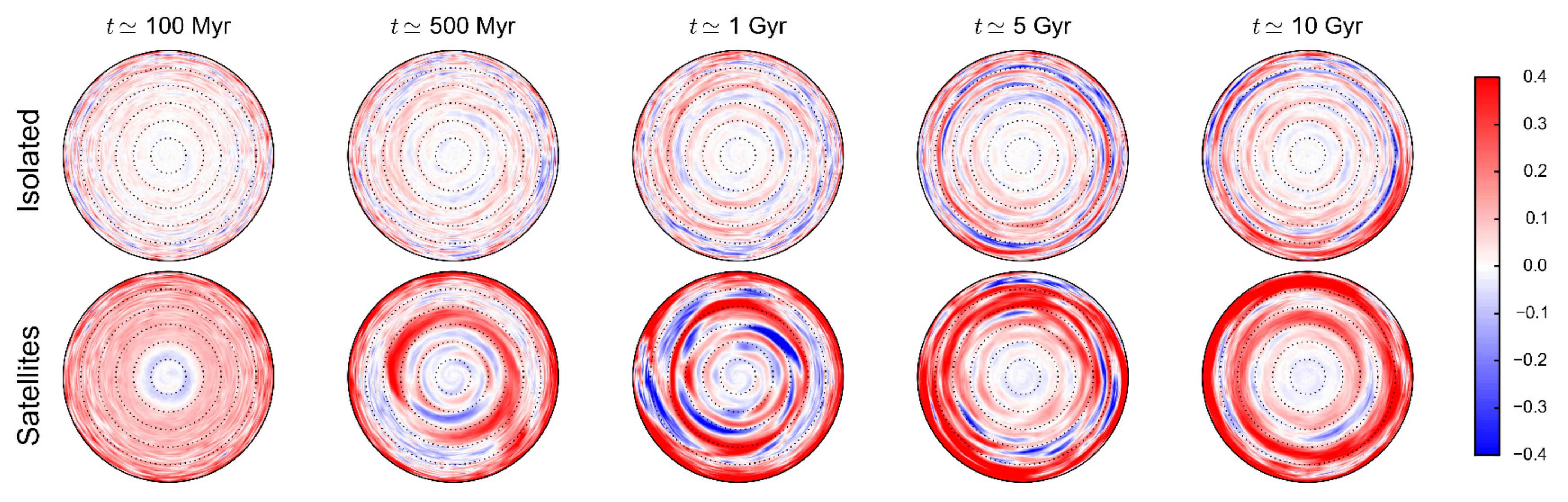}
\caption{Surface overdensity for the control (top row) and satellite
  (bottom row) simulations at five epochs.  The colour map indicates
  the logarithm (base 10) of the ratio of the local surface density at
  the time indicated to the azimuthally-average surface density of the
  initial conditions.  Dotted concentric circles indicate increments
  of 5 kpc in radius. The rotation of the discs is counter-clockwise.
\label{fig:surfdensevo}}
\end{figure*}

In Fig.~\ref{fig:surfdensevo} we show the normalized surface density  
in the disc, $\Sigma\left(R,\,\phi,\,t\right )/\Sigma\left  
(R,\,t_0\right )$, where $\Sigma\left (R,\,t_0\right )$ is the initial  
surface density profile (see top panel Fig.~\ref{fig:discICs_radial}).  
The normalization is chosen to highlight non-axisymmetric features that
develop across the full range in $R$. Evidently, neither disc forms a
bar.  The disc in the control simulation develops tightly wrapped
flocculent spiral arms with a density contrast that tends to be more
pronounced at late times and toward the outer region of the disc.  The
spiral structure is considerably stronger in the satellite simulation.
At early times the density features are more asymmetric and irregular.
Note, in particular, what appears to be evidence of a single satellite
interaction in the $t=500\,{\rm Myr}$ snapshot at $R\sim 20 - 25\,{\rm
  kpc}$ and roughly the `5-o'clock' position of the disc.  At later
times large density contrasts survive only past $R \sim 10 - 15 \,
{\rm kpc}$ and extend in azimuthal angle around a large fraction of
the disc.  By the end of the simulation the disc develops a dominant
$m = 1$ spiral arm that is morphologically leading.  We attribute
these density enhancements and morphological differences to collisions
or torques imparted from the orbiting subhalo system since they are
not present in the control run \citep{gauthier2006,dubinski2008}.

\begin{figure*} 
\includegraphics[width=\textwidth]{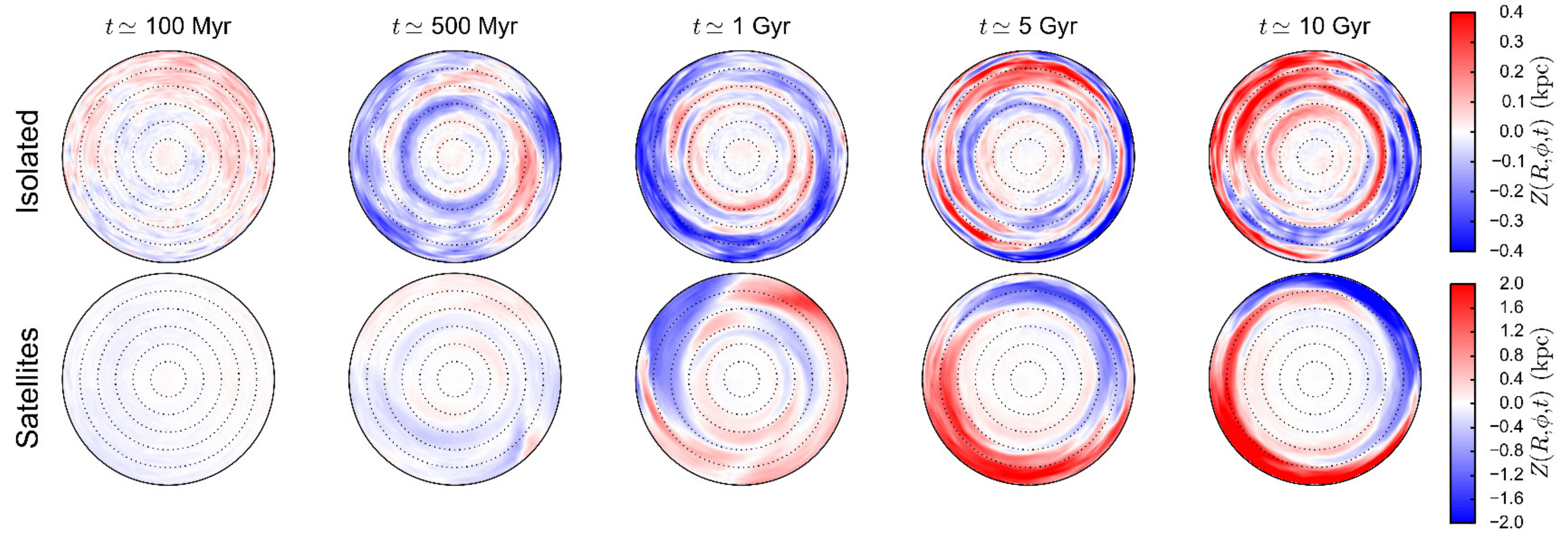}
\caption{Face-on maps of mean vertical displacement $Z(R, \phi, t)$
  for the isolated (top row) and satellite (bottom row) simulations at the
  same five epochs in Fig.~\ref{fig:surfdensevo}. Dotted concentric
  circles indicate increments of 5 kpc in radius. The rotation of the
  discs is counter-clockwise.
\label{fig:Zmapevo}}
\end{figure*}

In Fig.~\ref{fig:Zmapevo} we show vertical displacement, $Z\left
(R,\,\phi,\,t\right )$, across the disc at the same epochs as those
shown in Fig.~\ref{fig:surfdensevo}. The trends are similar to those seen in the surface density plot.  Bending waves appear in both the
control and satellite simulations though they have a higher amplitude
by a factor of $\sim 5$ in the satellite case.  In both cases, strength in the bending waves tends to
migrate toward the edge of the disc over time.

At early times, there are localized bending waves in the satellite case.  In particular, we see evidence of an apparent interaction event in the $500\,{\rm Myr}$ snapshot at the same location in the disc as seen in the surface density, and have identified the subhalo responsible for this event. This satellite plunged through the disc midplane at a radius of $\sim$~26.5 kpc at $\sim$~400 Myr (indicated by the star symbol in Fig.~\ref{fig:satimpact}) on a prograde orbit with cylindrical velocities $(V_R , V_\phi , V_z) \simeq (-160, \, 50, \, 210) \, {\rm km} \, {\rm s}^{-1}$. At the time of impact the satellite had a mass of $\sim 8 \times 10^{8} \, M_{\odot}$ within its 2.7 kpc radius Jacobi sphere. Despite crossing through the disc with such a large vertical velocity, the radial extent, surface density, and in-plane velocity of the satellite are significant, and therefore it is not surprising that the impact left such a clear and spatially extended mark in the disc, as seen in Figs.~\ref{fig:surfdensevo} and \ref{fig:Zmapevo}.

We next consider the application of Fourier methods to the
simulations.  Formally, the DF of an $N$-body system can be expressed
as the sum of six-dimensional $\delta$-functions centred on the phase
space coordinates (positions and velocities) of the particles and
weighted by their mass.  The density is found by integrating over
velocities while the surface density is found by further integrating
over $z$.  Thus, the surface density is given by
\begin{equation}
\Sigma\left (R,\,\phi,\,t\right ) = \sum_j m_j \, \delta \left (\phi -
\phi_j(t)\right ) \frac{\delta \left ( R-R_j(t)\right )}{R}~.
\end{equation}
\noindent (For the maps in Fig.~\ref{fig:surfdensevo} we effectively
integrate over bins in a polar grid and divide by the area of each
bin.  In Fig.~\ref{fig:Zmapevo}, we compute the average $z$ in each of
these bins).
\\
\\
Beginning with the pioneering work of \citet{sellwoodathanassoula1986}
it has proved useful to consider various transforms of the surface
density when studying the formation of warps, bars, and spiral
structure.  A Fourier series in $\phi$ yields a decomposition of
$\Sigma$ in terms of azimuthal wave number $m$, with warps dominated
by $m=1$ while bars and bisymmetric spiral structure are
dominated by $m=2$.  A Fourier transform in $t$ yields the
surface density amplitude as a function of $\omega$, $R$, and $\phi$,
or $\omega$, $R$, and $m$.  The frequency $\omega$ can be used as a
proxy for angular pattern speed.  \citet{sellwoodathanassoula1986}
also consider a Fourier transform in $\ln{R}$.

In what follows, we apply Fourier methods to bending waves.
We begin by dividing the disc into cencentric rings each
labelled by $\alpha$ and centred on a radius $R_\alpha$ with width
$\Delta R_\alpha$ and surface area $S_\alpha\equiv 2\pi R_\alpha
\Delta R_\alpha$.  The radially-smoothed surface
density is then
\begin{equation}
\Sigma\left (R_\alpha, \, \phi,\,t\right ) = 
2\pi \, S_\alpha^{-1}\sum_{j\in \alpha} m_j \, \delta\left (\phi - \phi_j(t)\right )~,
\label{eq:surfdens}
\end{equation}
\noindent where $R_\alpha \rightarrow R$ in the limit of small $\Delta
R_\alpha$ and the sum is over all particles in ring $\alpha$. The
vertical displacement, $Z\left (R,\,\phi,\,t\right )$, and vertical
bulk motion, $V_z\left (R,\,\phi,\,t\right )$, can be constructed by
taking moments of the DF.  In analogy with
equation~(\ref{eq:surfdens}) we have
\begin{equation}
Q \left (R_\alpha \, \phi,\,t\right ) = 
2\pi \, S_\alpha^{-1} \, \Sigma^{-1}\left (R_\alpha, \, \phi,\,t\right )
\sum_{j\in \alpha} m_j \, q_j(t) \, \delta\left (\phi - \phi_j(t)\right )~,
\end{equation}
\noindent where $Q = \left\{Z,\,V_z,\,\dots \right\} $ and $q_j = \left
\{ z_j,\,v_{z,j},\,\dots\right\}$, and the $\dots$ refer to any
non-linear function of $z$ and $v_z$.

The Fourier series for the surface density is given by
\begin{equation}
\Sigma\left (R_\alpha, \, \phi,\,t\right ) = \sum_{m=0}^\infty
\Sigma_m (R_\alpha, \, t) e^{-im\phi}~,
\end{equation}
\noindent where
\begin{equation}
\begin{split}
\Sigma_m (R_\alpha, \, t) & = \frac{1}{2\pi} \int_0^{2\pi} \Sigma \left (R_\alpha, \, \phi,\,t\right )
e^{im\phi} \, d \phi\\
& = S_\alpha^{-1}\sum_{j\in \alpha} m_j e^{im\phi_j}~.
\end{split}
\end{equation}
Likewise, the Fourier coefficients of $Q \left (R_\alpha, \, \phi,\,t\right )$ 
are 
\begin{equation}
Q_m (R_\alpha, \, t) = S_\alpha^{-1} \sum_{j\in\alpha}
\frac{m_j \, q_j \, e^{-im\phi_j}}{\Sigma\left (R_j,\,\phi_j,\,t\right )}~,
\end{equation}
\noindent where $\Sigma\left (R_j,\,\phi_j,\,t\right )$ is the
interpolated surface density at the position of the $j$th particle.

\begin{figure*} 
\includegraphics[width=\textwidth]{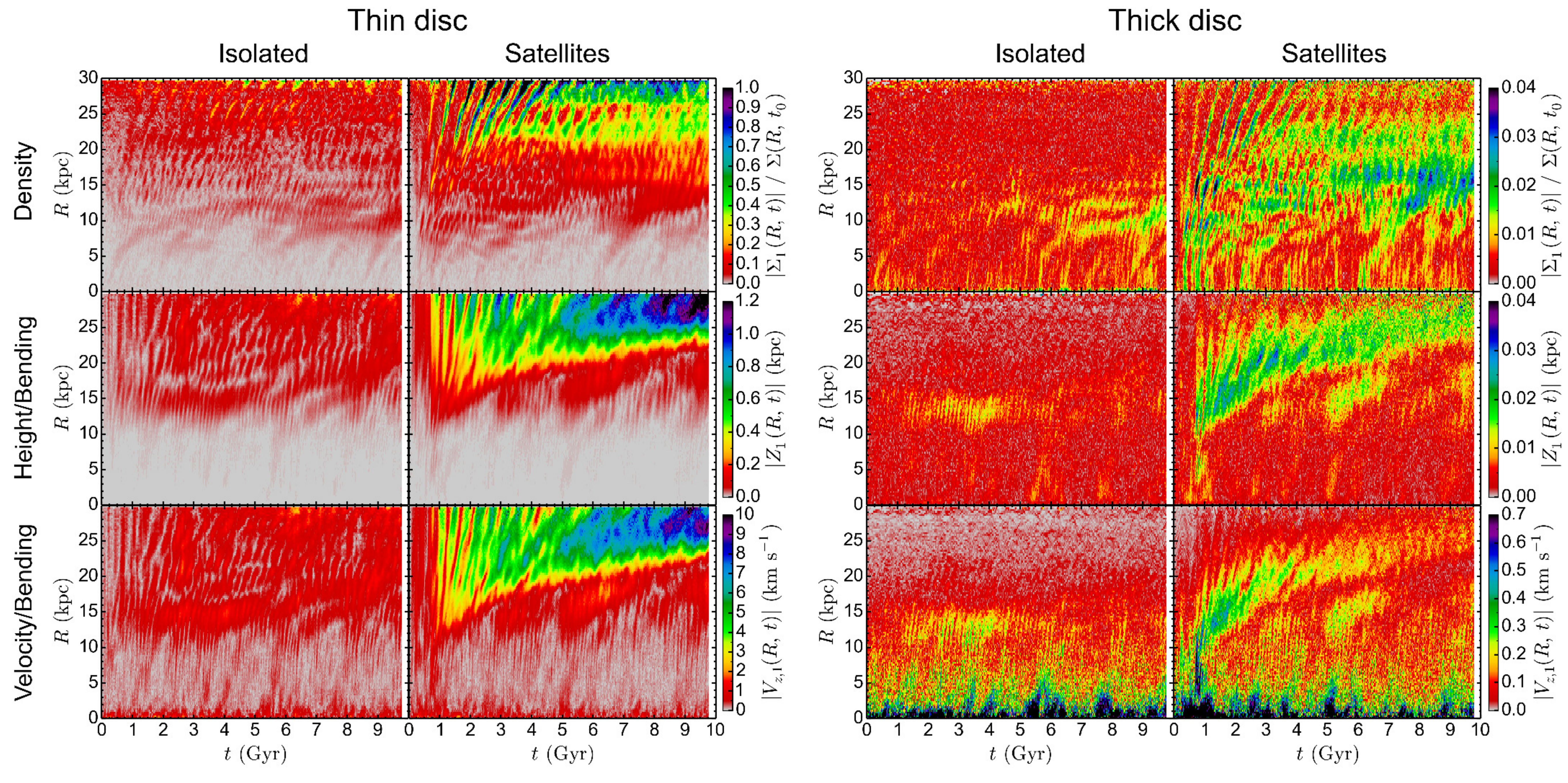}
\caption{Time evolution of $m=1$ Fourier coefficients for in-plane density
  and bending waves in both the thin and thick discs of the isolated
  and satellite simulations, as indicated. Note the difference in colour
  scale between the thin and thick discs.
\label{fig:m1coefstime}}
\end{figure*}

In Fig.~\ref{fig:m1coefstime} we plot
$|\Sigma_1 (R, \, t)| \, / \, \Sigma (R, \, t_0)$, $|Z_1 (R, \, t)|$,
and $|V_{z,1} (R,\,t)|$. In particular, we show the separate contributions
from the thin and thick discs, and compare the isolated and satellite
simulations.  Generally, we see that the thick disc is much less
responsive by a factor of $4\sim 5$ and the subhaloes induce
perturbations with peak amplitudes $\sim 5$ times greater than those
in the control simulation.

Bending waves are strongest in the outer disc for both control and
satellite runs. This result is expected for several reasons.
First, the transition from a relatively quiet disc to one with
stronger vertical waves occurs at $R \sim 15 {\rm kpc}$, which roughly
coincides with region where the surface density of the thick disc
falls to negligible values. Thus, the increased gravity from a
relatively warm component may make the inner disc less responsive to bending perturbations.  Moreover, the higher surface density of the
thin disc itself acts to stiffen the disc.  The outer disc is
dominated by the cooler thin disc though much of the gravitational
field is from the halo, and it is therefore more prone to
non-axisymmetric structure formation and bending waves.  Finally,
as noted in section~\ref{sec:subhalointeractions} the strongest
subhalo encounters occur in the outer disc.

\begin{figure*} 
\includegraphics[width=\textwidth]{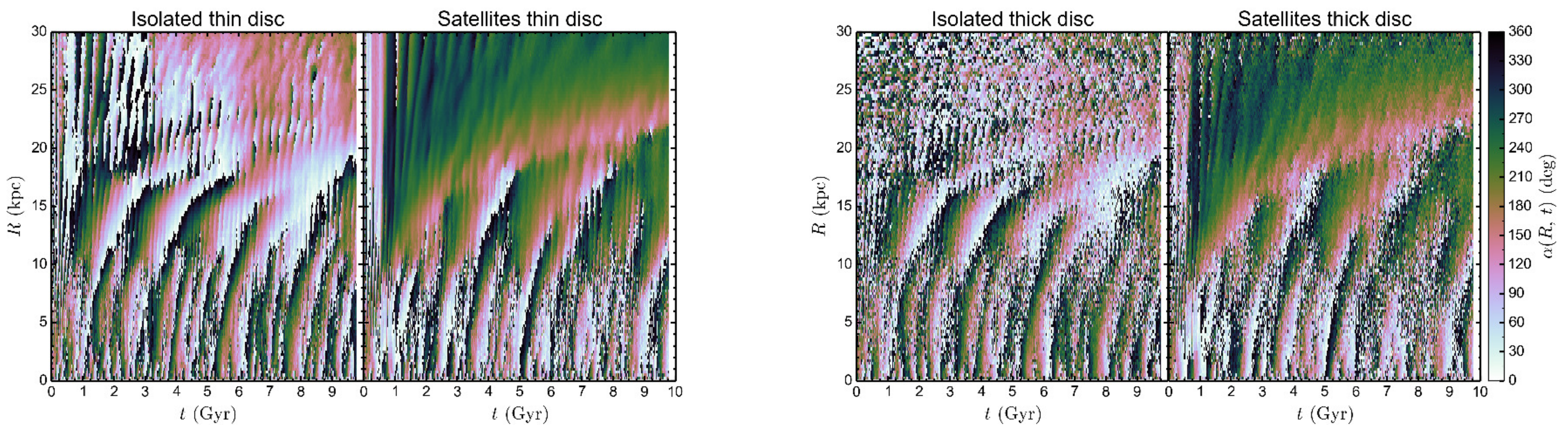}
\caption{Time evolution of Fourier phase angle $\alpha (R, \, t)$ for $m=1$ vertical displacement bending waves in the thin and thick discs of both simulations, as indicated.
\label{fig:m1coefphasetime}}
\end{figure*}

To determine the morphology of $m=1$ bending waves we compute
the complex phase for $Z_1$,
\begin{equation}
\alpha (R, \, t) = {\rm tan}^{-1} \Bigg[ \frac{{\rm Im} \{Z_1 (R, \,
    t) \} }{{\rm Re} \{ Z_1 (R, \, t) \} } \Bigg]~,
\end{equation}
\noindent as a function of $R$ and $t$.  In
Fig.~\ref{fig:m1coefphasetime} we show $\alpha (R, \, t)$ for the thin
and thick discs in both simulations. Attention is immediately drawn to
ridges of constant phase that arc outward in radius over time. Since
the coordinate system in our simulations has the disc rotating
counter-clockwise (i.e., in the direction of increasing $\alpha$), we
conclude that the bending waves are either prograde and trailing or
retrograde and leading. A distinction between the two cases is made
when considering the behaviour of $\alpha$ over slices in $R$ and
time. The signature of leading waves is for $\alpha$ to increase with
increasing $R$ at fixed time.  On the other hand, the signature for a
retrograde bending wave is for $\alpha$ to decrease with time at fixed
$R$. Indeed, we observe these very relationships in
Fig.~\ref{fig:m1coefphasetime} and conclude that the dominant
long-lived bending waves in our simulations are morphologically
leading with patterns that rotate in a retrograde sense, as is expected for bending waves (see section 8.1 of \citealt{sellwood2013}).  We note that
though we only show $\alpha$ for vertical displacements the phase
angle for bulk vertical motions displays the same behaviour as a
function of $R$ and $t$, albeit with an offset.

\section{Spectral wave Analysis}
\label{sec:spectralwaveanalysis}

\begin{figure*} 
\includegraphics[width=\textwidth]{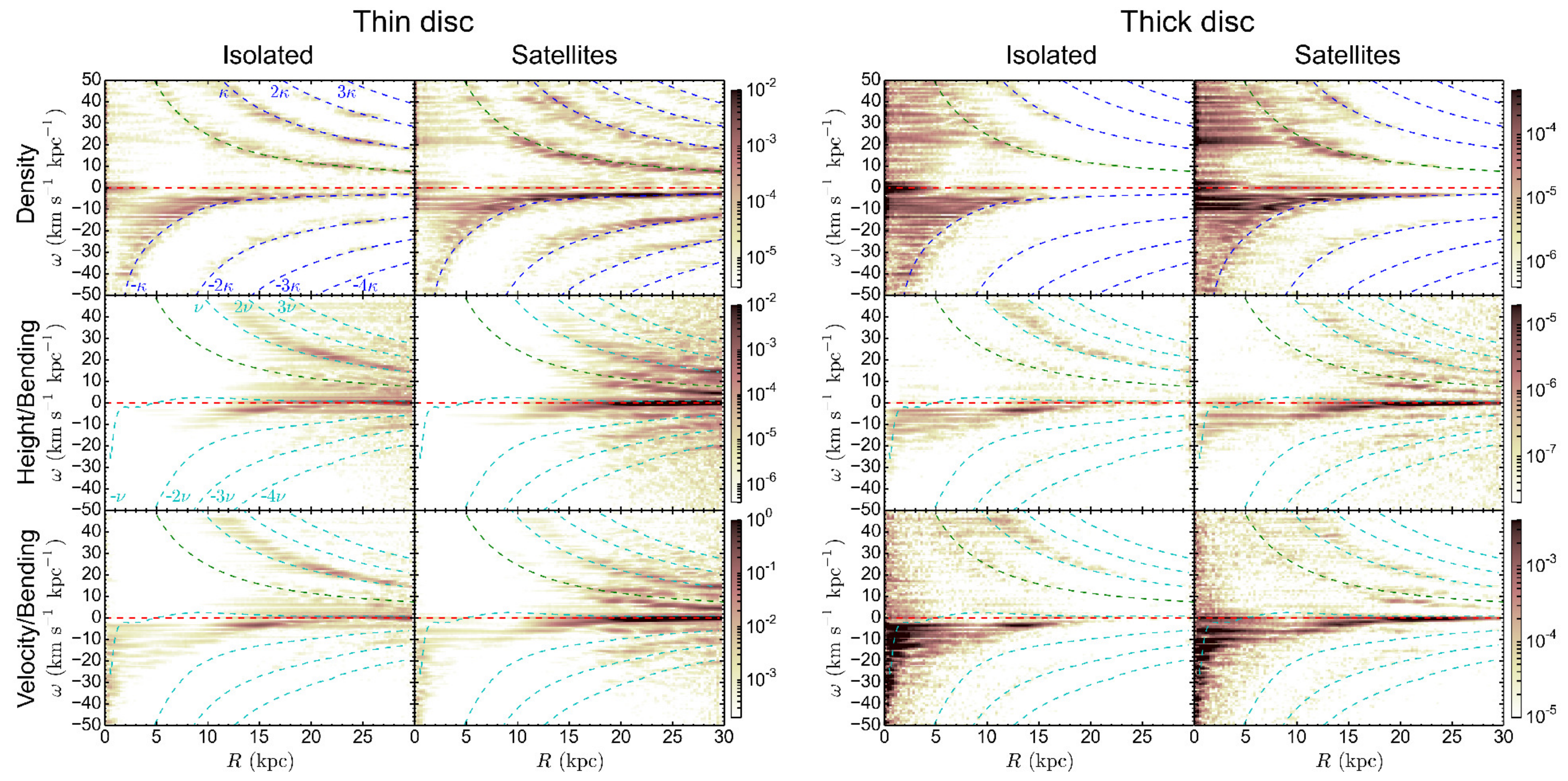}
\caption{Frequency power spectra for $m=1$ in-plane density and
  bending waves in the thin and thick discs of both simulations over a
  time baseline of $0 \le t \lesssim 10 \, {\rm Gyr}$. The layout of
  the figure is analogous to Fig.~\ref{fig:m1coefstime}. The
  horizontal dashed red line references zero frequency while the
  dashed green curve corresponds to corotation. Overlaid as dashed
  blue and cyan curves are orbital resonances for kinematic density
  waves in a cold disc (see equation~\ref{eq:orbitresonances}), as
  indicated, corresponding to the initial conditions.
\label{fig:m1coefsfreq}}
\end{figure*}

The spectral methods developed by \citet{sellwoodathanassoula1986}
allow one to characterize density waves in terms of angular frequency
and galactocentric radius. These methods provide valuable insight into
disc dynamics and have been extended to the study of bending waves
\citep[see][]{chequers2017}.

Our spectral analysis for in-plane density waves and bending waves
begins with the Fourier coefficients, $\Sigma_m (R,\,t_j)$ and $Q_m
(R,\,t_j)$, for $N$ snapshots at times $t_j=j \, \Delta_t +t_0$, where
$\Delta_t$ is the time between snapshots, $j=0\dots N-1$, and $N$ is
even (see Fig.~\ref{fig:m1coefstime}). We then perform a discrete
Fourier transform on the time series to obtain the two-sided frequency
coefficients \citep[][Section 13.4]{press2007} as
\begin{equation} \label{eq:freqfouriercoefs}
	F_m (R,\, \omega_k) = \begin{cases}
		\Sigma^{-1} (R,\,t_0) \sum_{j = 0}^{N-1} \Sigma_m (R, t_j) \, w(j) \, e^{2 \pi i j k/N}, \\
		\sum_{j = 0}^{N-1} Q_m (R, t_j) \, w(j) \, e^{2 \pi i j k/N}~. 
	\end{cases}
\end{equation}
\noindent Here, $w(j)$ is a Gaussian window function with a standard
deviation of $N/2^{5/2}$, which is introduced to diminish
high-frequency spectral leakage. The discrete frequencies are given by
\begin{equation}
\omega_k = \frac{2 \pi}{m} \frac{k}{N \Delta_t}~,
\end{equation}
\noindent with $k = -N/2\dots N/2$. The frequency resolution is
regulated by the length of the time baseline, $N \Delta_t$, while the
Nyquist frequency, corresponding to $\omega_{k= \pm N/2}$, is derived
from the time resolution, $\Delta_t$. The frequency power spectrum is
then computed as
\begin{equation}
P_m (R,\, \omega_k) = \frac{1}{W} \, |F_{m} (R, \,\omega_k)|^2 ~,
\end{equation}

\noindent where $W = N \, \sum_{j = 0}^{N-1} w(j)$ is the window
function normalization.

In Fig.~\ref{fig:m1coefsfreq} we plot the frequency power spectra for
$m=1$ density and bending waves over a time baseline of $0 \le t
\lesssim 10 \, {\rm \, Gyr}$. The general properties of relative
frequency power between the isolated and satellite runs as well as the
thin and thick discs is analogous to that of the time series of
Fourier coefficient magnitude in Fig.~\ref{fig:m1coefstime} (Note the
different scales used for power between the thin and thick discs).
    
Evidently, bending wave power follows one of a series of branches that arc
across the $R$-$\omega$ plane and is concentrated in the outer disc at
lower, or even counter-rotating, frequencies. These observations are
in accord with our analysis in Section~\ref{sec:fourierwaveanalysis}
where we concluded that the dominant waves in the disc were
morphologically leading and propagate in a retrograde fashion compared
to the rotation of the disc.

The power spectrum branches in Fig.~\ref{fig:m1coefsfreq} 
roughly coincide with the orbital resonance curves
\begin{equation} \label{eq:orbitresonances}
  \omega\,(R) = \begin{cases}
    \Omega \, (R) + n \, \kappa\,(R), & \text{$m=1$ in-plane density waves},\\
    \Omega \, (R) + n \, \nu \, (R), & \text{$m=1$ bending waves}~. 
  \end{cases}
\end{equation}
\noindent Here, $\Omega$ is the circular frequency, $\kappa$ is the total
epicyclic frequency, and $\nu$ is the total vertical forcing frequency
from the disc, bulge, and halo.  The integer $n$ corresponds to the number of
azimuthal periods an orbit takes to close during one radial
oscillation period in a frame rotating with angular frequency
$\omega$.  The Lindblad resonances and their vertical analogs
correspond to $n = \pm 1$.

\begin{figure*} 
\includegraphics[width=\textwidth]{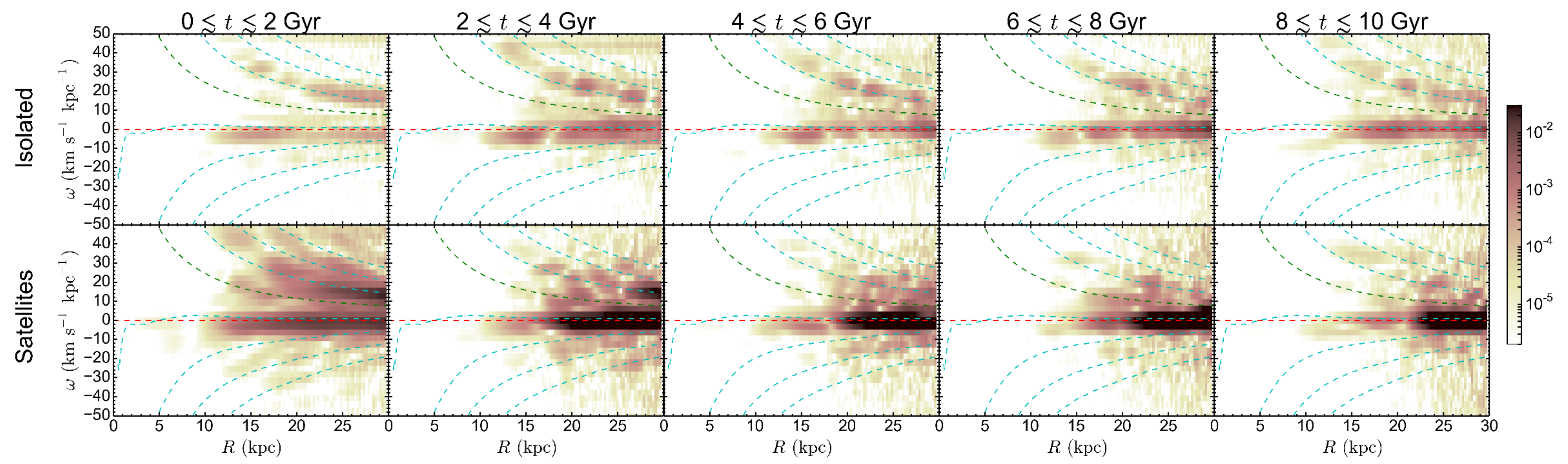}
\caption{Time evolution of $m=1$ height/bending frequency power
  spectra evaluated over $\sim$~2 Gyr intervals for the thin disc in
  both simulations, as indicated. The dashed curves correspond to the
  same quantities as in Fig.~\ref{fig:m1coefsfreq}.
\label{fig:m1coefsfreq_timeseries}}
\end{figure*}

Many of the features in Fig.~\ref{fig:m1coefsfreq} can be understood 
by appealing to the WKB approximation, where one assumes that 
the waves (either density or bending) are tightly wound.  In this 
approximation, and under the further assumption that the disc is cold 
and thin, the dispersion relations are 
\citep[][sections 6.2.1 and 6.6.1, and section 8.1 from \citealt{sellwood2013}]{binney2008}
\begin{equation} \label{eq:disprelation}
(\omega - \Omega)^2 = \begin{cases}
	\kappa^2 - 4\pi^2 G \Sigma / \lambda_{\rm d}, & \text{for $m=1$ density waves}, \\
	\nu^{2} + 4\pi^2 G \Sigma / \lambda_{\rm b}, & \text{for $m=1$ bending waves,}
  \end{cases}
\end{equation}
\noindent 
where $\Sigma$ is the radial surface density and $\lambda_{\rm d}$
$(\lambda_{\rm b})$ is the wavelength of a density (bending) wave. From 
equation~(\ref{eq:disprelation}) we see that 
WKB density waves are `forbidden' in the region outside the Lindblad 
resonances while bending waves are forbidden between the corresponding 
vertical resonances (equation~\,\ref{eq:orbitresonances}).  Indeed, we 
find that in the case of density waves power mainly lies between the Lindblad 
resonance curves whereas power lies outside the $n = \pm 1$ vertical 
resonances in the case of bending waves. 

The Lindblad resonances and their vertical analogs emerge from
equation~(\ref{eq:disprelation}) in the limit that the surface density terms (i.e. the terms that encode the forcing frequency due to the self-gravity of the perturbation)
on the right hand side are negligible when compared to the $\kappa^2$
or $\nu^2$ terms.  Thus, we can predict the degree to which power
should align with the resonance curves by comparing the two terms on
the right-hand side of equation~(\ref{eq:disprelation}).  In general, we
expect this alignment to be better at large radii since the surface
density falls off exponentially while $\kappa$ and $\nu$ fall as
as powers of $R$.  Furthermore, following similar arguments to those found in
\citet[][equation 3.89]{binney2008}, one can show that
\begin{equation}\label{eq:nukap}
\frac{\nu^2}{\kappa^2} \simeq \frac{3}{2}\frac{\rho}{\bar{\rho}}
\end{equation}
\noindent where $\bar{\rho}$ is the mean density within a given radius
and $\rho$ is the density of the galaxy in the midplane of the disc at that radius. Since the bulge and halo are nearly spherically symmetric with monotonically decreasing density profiles and the submaximal disc lies in the plane, we conclude that $\nu < \kappa$. Therefore the surface density term in
equation~(\ref{eq:disprelation}) will play a larger role for bending waves
than for density waves. It is therefore not surprising that bands in
power are more closely aligned with the resonance curves in the
density wave case than the bending wave one.

In Fig.~\ref{fig:m1coefsfreq_timeseries} we show a time sequence of
$m=1$ bending waves over $\sim$~2 Gyr intervals for the thin disc.  For
each of the time intervals in the control simulation, power lies along
branches just outside the $n=\pm 1$ resonance curves, as expected.
The situation is more complicated in the case with satellites.  At
early times, there are numerous regions of power across the $R-\omega$
plane that are not particularly well correlated with the resonance
curves.  Over time, the positive frequencies diminish in power,
although localized waves still exist, and the strength of
low-frequency counter-rotating waves increases in strength and
migrates outward. By the end of the simulation, the branches of power
roughly mirror that of the isolated simulation, though the relative
weighting of the waves localized in radius is quite different.  These
results make intuitive sense.  Bending waves in the control simulation
involve the growth of linear perturbations, which are well described
by a WKB analysis.  The satellite-provoked waves, on the other hand,
are more random in nature, though over time, the most persistent 
ones also appear to coincide with predictions from linear theory.

\section{Correlation between $Z$ and $V_z$}
\label{sec:wavelikenature}

Simultaneous observations of bulk vertical motions and vertical
displacement provide an avenue toward understanding wave dynamics in
the Galactic disc.  Roughly speaking the angular frequency of a wave
will be given by the amplitude of its oscillations in velocity divided
by the amplitude of its displacement oscillations.  \citet{xu2015}
found evidence for corrugations in the disc with an amplitude of about
$120\,{\rm pc}$ and a wavelength of $7\,{\rm kpc}$ at Galactocentric
radii of $10-16\,{\rm kpc}$.  On the other hand, \citet{schonrich2017}
identified ripples in the velocity field of the Solar Neighbourhood
with an amplitude of $\sim0.8\,{\rm km\,s^{-1}}$ and a wavelength of
$2-2.5\,{\rm kpc}$.  Of course, the wavelength and amplitude of a
bending wave might change with Galactocentric radius. As noted
in \citet{schonrich2017}, the extent of their sample was not sufficient to
detect such changes.  In any case, if one assumes that the
\citet{xu2015} and \citet{schonrich2017} oscillations are related,
then the inferred angular frequency is $\simeq 6.7\,{\rm
  km\,s}^{-1}{\rm kpc}^{-1}$, which correponds to a period of about
$920\,{\rm Myr}$.  Obviously, the caveats mentioned above imply that this
is only a ballpark value and valid only if the two observations are of the same wave-like structure.

\begin{figure*} 
\includegraphics[width=\textwidth]{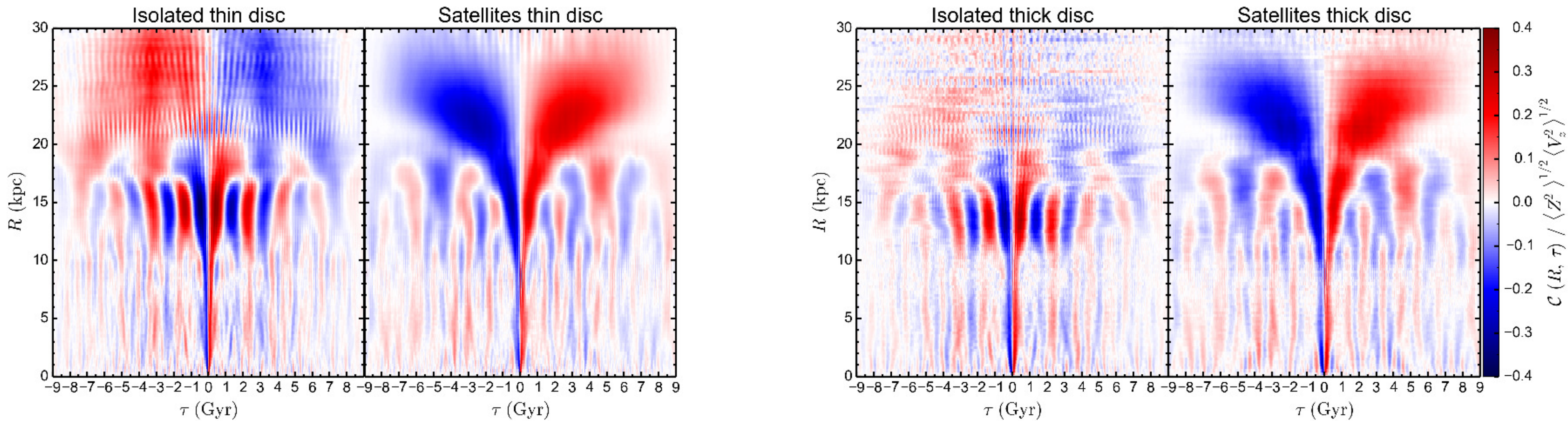}
\caption{Normalized cross correlation (see equations
  \ref{eq:crosscorr} and \ref{eq:crosscorrrms}) between $m=1$ velocity
  and displacement bending waves. The column titles indicate which
  disc and simulation each panel corresponds to.
\label{fig:wavecorrelation}}
\end{figure*}

The correlation between bulk vertical motion and midplane displacement in simulations
has been discussed by \citet{gomez2013}, \citet{gomez2016},
\citet{chequers2017}, and \citet{gomez2017}.  In both of our control
and satellite simulations the bulk vertical motions follow the same
general patterns as the displacements
(c.f. Figs.~\ref{fig:m1coefstime}-\ref{fig:m1coefsfreq}) except for a
phase offset.  This behaviour is expected for wave-like motion and
allows one to infer a rotational frequency from the respective
amplitudes of each pattern.

To further explore the connection between velocity and displacement
oscillations, we compute the azimuthally-averaged 
cross correlation function between $Z_1$ and $V_{z,1}$:

\begin{equation} \label{eq:crosscorr}
\mathcal{C}\,(R,\,\tau) = \frac{1}{2 \pi} \int_{0}^{2\pi} d\phi \,
\frac{1}{t_f} \int_{0}^{t_f} V_{z,1} (R, \, \phi, \, t) \, Z_1 (R, \,
\phi, \, t + \tau) \,dt~,
\end{equation}

\noindent where $\tau$ is the lag time between the signals. To
highlight small amplitude wave-like features we normalize to the
root-mean-square of $Z$ and $V_z$ at a given radius,

\begin{equation} \label{eq:crosscorrrms}
\langle Q^2 \rangle^{1/2} = \sqrt{ \frac{1}{t_f} \int_{0}^{t_f} q^{2} (t) \, dt}~,
\end{equation}

\noindent where $Q = \{Z, \, V_z\}$ and $q = \{z, \, v_z\}$, respectively.

For the highly idealized case of a linear one-dimensional
monochromatic plane wave, the displacement, $z = D \, {\rm sin} (\omega t + \phi)$, is related to the velocity by $v = d \, z /dt = D \, \omega \, {\rm cos} (\omega t + \phi)$. The displacement pattern will lag the velocity
by an offset in phase of $\pi/2$ and a time $\pi/2\omega$. Thus,
the cross correlation between the two waveforms, $\mathcal{C} \,
(\tau) \propto {\rm sin} (\omega \, \tau)$, will display periodic peaks at
$2p -1$ multiples of $\tau = \pi/2\omega$, where $p$ is an integer,
that decreases in amplitude with increasing $|\tau|$.

In Fig.~\ref{fig:wavecorrelation} we show the normalized cross
correlation between the $m=1$ displacement and velocity patterns in
our simulations and immediately notice large correlation amplitudes
that are localized in $R$, periodic in $\tau$, and differ
between the two simulations. We also note that the radii at which these
peaks in correlation amplitude occur seem to correspond to radii that
feature large frequency power as seen in the spectra in
Fig.~\ref{fig:m1coefsfreq}.

Of course, in the case of our simulations the situation is more
complex than simple one-dimensional monochromatic waves since our
discs comprise waves of varying frequency localized in radius. Quite
remarkably, for a given $R$ the cross correlation patterns we see in
Fig.~\ref{fig:wavecorrelation} generally exhibit the exact behaviour
that is expected for simple plane waves. In particular, we observe
fluctuations in amplitude that are periodic in $\tau$ and are able to
recover the localized frequency $\omega = \pi/2\tau$. More globally,
we see intricate patterns of correlation amplitude that highlight the
dominant long-lived waves themselves and the spatial transitions
between them.

We note that in the case of the thin disc in our isolated galaxy simulation for $R \gtrsim 20$ kpc the correlation amplitude is
negative and positive to the right and left of zero lag, respectively,
which is interpreted as the displacement pattern \textit{leading} the
velocity. This behaviour may indicate non-linear dynamics or damping.

\section{Discussion}
\label{sec:discussion}

There is now evidence for bending and breathing waves in the stellar disc
of the Milky Way from numerous astrometric surveys including GDR2.
Nevertheless, the origin of these waves remains uncertain.  One
attractive proposal is that they are excited by interactions between
the disc and satellite galaxies or dark matter subhaloes.  This
proposal is a natural extension of the hypothesis that satellite-disc
encounters heat and thicken the disc \citep{toth1992} and may also
prompt the formation of a bar or spiral structure
\citep{gauthier2006, dubinski2008, kazantzidis2008}.  In the simplest
  interpretation wave-like features in the disc can be associated with
  particular disc-satellite interactions.  For example,
  \citet{purcell2011} used $N$-body simulations to show that the passage
  of the Sagittarius dwarf galaxy through the disc may have been
  responsible for the Milky Way's bar and spiral structure.
  \citet{gomez2013} then showed that the same encounter could generate
  vertical waves similar to the ones seen in the data.

The numerical experiments by \citet{gauthier2006} and
  \citet{dubinski2008} paint a more complicated picture.  They
  simulated a disc galaxy with a clumpy halo and found that halo
  substructure vigourously excited spiral density waves.  Moreover,
  even though their equilibrium (i.e., smooth halo) model was stable
  to bar formation for at least 10 Gyr, subhaloes could trigger the formation of a bar --
  though the timing of this depended on the particulars of the subhalo
  distribution.  The implication is that the Milky Way's dynamical
  state may reflect both recent, singular encounters and the
  cumulative effect of many disc-satellite interactions over its
  lifetime.
 
  A related question concerns the evolution of perturbations once they
  are excited.  \citet{delavega2015} suggested that this evolution can
  be described as kinematic phase mixing and that the self-gravity of
  the waves can be ignored.  If true, this approximation would greatly
  simplify the analysis of perturbations in the disc.  Indeed,
  \citet{antoja2018} used this idea to interpret the spiral structure
  they found in ${\bf v}-z$ phase space projections of the GDR2 data.
  Simple kinematics allowed them to `unwind' these structures and
  hence `date' the event that gave rise to them.

  Of course, similar arguments, when applied to the problem of spiral
  structure, lead to the so-called winding problem wherein spiral arms
  become so tightly wound in just a few orbital periods that they
  would dissolve into the rest of the disc. In an attempt to resolve the winding problem it was proposed that spiral arms are density waves, which are amplified and maintained by self-gravity \citep[][but see also \citealt{lindblad1963}, \citealt{lin1966}, \citealt{lin1969}, and \citealt{toomre1969}]{lin1964}. Similar
  arguments have been made in the study of galactic warps.  For
  example, the eigenmode analyses of bending waves by
  \citet{hunter1969}, \citet{sparke1984}, and \citet{sparke1988} explicitly include
  self-gravity by treating the disc as a system of
  gravitationally-coupled concentric rings.  When one ring is displaced
  from its equilibrium position, it exerts a perturbing force on the
  unperturbed rings and also feels a restoring force due to them. It is the combination of these two effects that yields a continuum of bending waves, which could describe the waves seen by \citet{xu2015} and \citet{schonrich2017}.

The simulations described in this paper allow us to explore these
issues in more detail.  In particular, we compare the evolution of a
stellar-dynamical disc embedded in a clumpy dark halo with one that
evolves in a smooth halo.  In analyzing our simulations we turn to the
spectral techniques laid out in \citet{sellwoodathanassoula1986}, and
extended to bending waves by \citet{chequers2017}
(c.f. Fig.~\ref{fig:m1coefsfreq}).  We find that the dominant
long-lived waves for both in-plane and vertical disturbances with $m=1$ symmetries lie mainly on or
around two main branches in the $R$-$\omega$ plane, which roughly
coincide with the Lindblad resonances and their vertical
counterparts, and is in agreement with predictions from a linear eigenmode analysis of bending waves \citep[][but see \citealt{chequers2017}]{hunter1969,sparke1984,sparke1988}. However, here we make a distinction between `modes' of the disc and waves since the radial locations of the waves we observe in our simulations change over time (c.f Figs~\ref{fig:surfdensevo}-\ref{fig:m1coefphasetime}) and only some of the waves appear to have somewhat well-defined frequencies. \citet{chequers2017} showed that bending waves on the upper frequency branch are trailing and rotate prograde to that of the disc while the lower branch corresponds to retrograde leading waves.

Rather astonishingly, we find the dominant waves to lie on these two branches
for \textit{both} the isolated and satellite simulations. Therefore,
the long-lived waves that the subhalo collisions excite are not
haphazard, but are the very same ones that arise in the absence of
halo substructure. This implies that the location of waves on the $R$-$\omega$ plane is largely dictated by the structure of the disc, which is model dependent. The $m=1$ waves we observe in face-on maps
of the galaxies (i.e. Figs.~\ref{fig:surfdensevo} and
\ref{fig:Zmapevo}) manifest as a superposition of waves from the two frequency branches. Any differences in bending wave morphology between the galaxies is attributed to the relative weighting of the upper and lower branches.

\section{Conclusions}
\label{sec:conclusions}

In this paper, we explore the continual generation of bending waves by
a system of satellites or dark matter subhaloes.  (Though our focus
here is on bending waves, we note that higher order waves, such as
breathing waves, are also generated). In particular, we show that the
long-lived bending waves that arise when the halo is clumpy are
qualitatively similar to, though stronger in amplitude than, the waves
that arise when the halo is smooth. Of course, the degree to which bending is excited depends on the characteristics of the subhalo distribution. Therefore, the study of vertical waves could provide insight into the properties of the Milky Way's subhalo population.  

One of the main objectives of this paper is to develop analytic tools
to study wave-like perturbations of the stellar disc, including those
that might be observed by \textit{Gaia} (See, for example,
\citealt{gaiadr2disckin2018}).  Indeed, soon after the release of GDR2
\citet{antoja2018} presented evidence of intriguing spiral structures
in various phase space projections of the data.  In that paper, the
authors work under the assumption that the vertical waves are
kinematic and generated by a single disc-substructure encounter.
These assumptions allow them to `unwind' the structure and infer an
approximate epoch for when the perturbation arose.

Our simulations suggest a more complicated scenario.  Consider first a
single satellite that passes through the disc.  Initially, the
satellite transfers momentum to the disc stars thereby imprinting a
local perturbation on the disc.  An example can be found in
Figs.~\ref{fig:surfdensevo} and \ref{fig:Zmapevo}.  The perturbation
is then sheared into roughly circular arcs and begins to act as a
bending wave, which can propagate through the disc.  Over time, energy
in the wave migrates to the edge of the disc causing a warp.  Energy
is also dissipated into the random motions of the disc stars thus
heating and thickening the disc.  Thus, the waves observed today will
likely involve a superposition of kinematic waves from recent events
and longer lived self-gravitating waves, which evolve on timescales much longer than kinematic phase wrapping.

Much of our analysis is based on the Fourier techniques developed by
\citet{sellwoodathanassoula1986} for the study of in-plane density
waves, and later extended by \citet{chequers2017} to the case of
bending waves. The time series of bending waves in the $R-\omega$
plane (Fig.~\ref{fig:m1coefsfreq_timeseries}) supports our picture.
In particular, at early times in the satellite simulation power is
randomly distributed across the $R$-$\omega$ plane.  These
perturbations, which we interpret as a superposition of kinematic
waves, quickly damp and shear.  At late times, after most of the
satellite interactions in the inner disc have occurred, the power is
more closely aligned with the resonance curves indicating the existence
of more organized bending waves.

Unsurprisingly, subhalo encounters excite bending waves to a larger
degree than in the case of an isolated disc.  However, the two cases
display similar waves in that they tend to be tightly wound,
morphologically leading, and rotate in a retrograde fashion compared
to that of the disc, as is expected for bending waves \citep{sellwood2013}. Generally, the bending waves are dominant in the
outer disc, and in particular just outside the edge of the thick disc
($R \sim 15 \, {\rm kpc}$ for our Milky Way-like model).  We attribute
this to the increased (radial and vertical) velocity dispersion and
self-gravity in the inner disc that act to resist bending
\citep{debattista1999}.

A novel feature of our simulations is the inclusion of two dynamically
distinct disc components.  We find that vertical bending waves are
significantly weaker in the kinematically hotter thick disc than
the thin one.  This observation has possible ramifications for
dynamical studies of the Milky Way that assume vertical
equilibrium. For example, the difference in response to compression
and rarefaction (breathing wave) perturbations between the thin and
thick discs could lead to systematic errors in an Oort-type analysis
\citep[e.g. ][]{bovy2013} for the vertical force and local surface
density (see \citealt{banik2017}).

The bending waves in our simulations are realized as spatial vertical
displacements as well as bulk vertical motions with a phase offset
\citep{gomez2013,gomez2016,chequers2017,gomez2017}.  We present a
cross-correlation analysis between these two manifestations of bending
and find that the waves are well described by simple plane waves, and
therefore their rotational frequencies can be easily inferred from the
amplitudes of spatial and velocity fluctuations as seen, for example, in
\citet{xu2015} and \citet{schonrich2017}.

In summary, our simulations suggest that bending of the Milky Way's stellar disc can be understood as a superposition of waves, which lend themselves to various analysis tools such
as linear theory, the WKB approximation, and spectral methods.  The
challenge is then to link theory and simulations of vertical waves
with observations.  The promise of \textit{Gaia} and future projects
such as the Large Synoptic Survey Telescope is that we will be able to
make this linkage and gain a better understanding of the Milky Way's
disc and the environment in which it lives.

\section*{Acknowledgements}

The authors would like to thank the referee for their insightful comments that improved the quality of this manuscript. The authors are also grateful to Jo Bovy and Alice Quillen for useful conversations, Jacob
Bauer for his assistance with the simulations, and Eugene Vasiliev for
his help with an early version of \textsc{agama} and comments on the original manuscript. We also acknowledge
the use of computational resources at Compute Canada and the Centre
for Advanced Computing. This work was supported by the Natural
Sciences and Engineering Research Council of Canada through the
Postgraduate Scholarship Program (MHC) and Discovery Grant Program
(LMW).




\bibliographystyle{mnras}
\bibliography{bibliography}

\newpage
\appendix

\section{Model parameters}

In Table~\ref{tab:initialmodelparams} we present the \textsc{agama}
model parameters used to construct the initial conditions of our Milky
Way-like disc.

\begin{table*}
\caption{The most sensitive \textsc{agama} input file parameters for each component's distribution function.}
\label{tab:initialmodelparams}
\begin{tabular}{lll}
\hline
\hline
Component & Parameter & Value\\
\hline
Thin disc &  Central surface density $\Sigma_{0}$ [$M_\odot \, {\rm pc}^{-2}$]  & 326  \\
&  Exponential scale radius $R_{d}$ [kpc]   &  3.6  \\
&  Central radial velocity dispersion $\sigma_{R,0}$ [${\rm km}\,{\rm s}^{-1}$]   & 103   \\
&  Central vertical velocity dispersion $\sigma_{z,0}$ [${\rm km}\,{\rm s}^{-1}$]   &  90  \\
&  Radial velocity dispersion exponential scale radius $R_{\sigma_R}$ [kpc]   &  7.2  \\
&  Vertical velocity dispersion exponential scale radius $R_{\sigma_z}$ [kpc] \, \, \, \,   &  7.2  \\
Thick disc &  Central surface density $\Sigma_{0}$ [$M_\odot \, {\rm pc}^{-2}$]  &  897  \\
&  Exponential scale radius $R_{d}$ [kpc]   &  2.01  \\
&  Central radial velocity dispersion $\sigma_{R,0}$ [${\rm km}\,{\rm s}^{-1}$]   &  155  \\
&  Central vertical velocity dispersion $\sigma_{z,0}$ [${\rm km}\,{\rm s}^{-1}$]   &  171  \\
&  Radial velocity dispersion exponential scale radius $R_{\sigma_R}$ [kpc]   &  5.5  \\
&  Vertical velocity dispersion exponential scale radius $R_{\sigma_z}$ [kpc] \, \, \, \,  &  7  \\
Bulge    & Mass normalization $M_0$ [$M_\odot$]    &  8.2 $\times$ $10^{10}$  \\
&  Break action $J_0$ [${\rm kpc}\,{\rm km}\,{\rm s}^{-1}$]   &  380  \\
&  Inner power law slope $\Gamma$   & 1.5   \\
&  Outer power law slope B  &  5   \\
&  Inner radial action velocity anisotropy coefficient $h_r$   &  1.2  \\
&  Inner vertical action velocity anisotropy coefficient $h_z$   &  0.9  \\
&  Outer radial action velocity anisotropy coefficient $g_r$  &  1.25  \\
&  Outer vertical action velocity anisotropy coefficient $g_z$   &  0.9   \\
&  Suppression action $J_{\rm max}$ [${\rm kpc}\,{\rm km}\,{\rm s}^{-1}$]  &  2800  \\
Halo    & Mass normalization $M_0$ [$M_\odot$]   &  5.4 $\times$ $10^{12}$   \\
&  Break action $J_0$ [${\rm kpc}\,{\rm km}\,{\rm s}^{-1}$]   & 3.3 $\times$ $10^{4}$   \\
&  Inner power law slope $\Gamma$   &  1.5  \\
&  Outer power law slope B  &  3.3  \\
&  Inner radial action velocity anisotropy coefficient $h_r$   & 1.4   \\
&  Inner vertical action velocity anisotropy coefficient $h_z$   & 0.7   \\
&  Outer radial action velocity anisotropy coefficient $g_r$  &  1.15  \\
&  Outer vertical action velocity anisotropy coefficient $g_z$   &  0.9   \\
&  Suppression action $J_{\rm max}$ [${\rm kpc}\,{\rm km}\,{\rm s}^{-1}$]  &  8.2 $\times$ $10^{4}$  \\
\hline
\hline
\end{tabular}
\end{table*}

\bsp	
\label{lastpage}
\end{document}